\documentclass[aip,jcp,reprint,superscriptaddress]{revtex4-1}

\usepackage{amsmath} 
\usepackage{graphicx} 
\usepackage{float}

\usepackage{soul}
\usepackage[usenames, dvipsnames]{xcolor}

\newcommand{\figurewidth}{8.5cm}
\newcommand{\narrowfigurewidth}{7.5cm}

\renewcommand{\vec}{\bf}

\begin{document}

\title{Gaussian approximation potential modeling of 
       lithium intercalation in carbon nanostructures}
\author{So Fujikake}
\affiliation{Engineering Laboratory, University of Cambridge,
             Cambridge CB2 1PZ, United Kingdom}
\affiliation{\'E{}cole des Ponts ParisTech, F-77455 Marne-la-Vall\'e{}e Cedex 2, France}
\affiliation{Department of Materials Engineering, The University of Tokyo, 7-3-1 Hongo, 
             Bunkyo-ku, Tokyo 113-8656, Japan}
\author{Volker L. Deringer}
\email{vld24@cam.ac.uk}
\affiliation{Engineering Laboratory, University of Cambridge,
             Cambridge CB2 1PZ, United Kingdom}
\affiliation{Department of Chemistry, University of Cambridge,
             Cambridge CB2 1EW, United Kingdom}
\author{Tae Hoon Lee}
\author{Marcin Krynski}
\author{Stephen R. Elliott}
\affiliation{Department of Chemistry, University of Cambridge,
             Cambridge CB2 1EW, United Kingdom}
\author{G\'a{}bor Cs\'a{}nyi}
\affiliation{Engineering Laboratory, University of Cambridge,
             Cambridge CB2 1PZ, United Kingdom}

\begin{abstract}
  We demonstrate how machine-learning based interatomic potentials can be used to
  model guest atoms in host structures. Specifically, we
  generate Gaussian approximation potential (GAP) models for the
  interaction of lithium atoms with graphene, graphite, and
  disordered carbon nanostructures, based on reference 
  density-functional theory (DFT) data.
  Rather than treating the full Li--C system, we demonstrate how
  the energy and force {\em differences} arising from Li intercalation
  can be modeled and then added to a (prexisting and unmodified) GAP model 
  of pure elemental carbon.
  Furthermore, we show the benefit of 
  using an explicit pair potential fit to capture ``effective'' Li--Li 
  interactions, to improve the performance of the GAP model.
  This provides proof-of-concept for
  modeling guest atoms in host frameworks with machine-learning based
  potentials, and in the longer run is promising for carrying out detailed
  atomistic studies of battery materials.
\end{abstract}

\maketitle

\section{Introduction}

Understanding and controlling the atomistic processes during charging
and discharging of batteries is a key requirement for developing
next-generation energy-storage solutions.
Among the most abundant technologies today are lithium (Li) ion batteries
in which the cathode is typically a complex oxide, whereas the anode is most 
commonly made of graphite or other carbonaceous nanostructures. 
\cite{Wu2003, Kaskhedikar2009, Etacheri2011}
Intercalation mechanisms and reactivity of Li in these materials 
have been widely studied using a range of experimental techniques.
\cite{Nishimura2008, Balke2010, Pecher2016}

In today's battery-materials research, first-principles computations
are routinely
used to complement experiments and even to make predictions,
usually based on density-functional theory (DFT). 
\cite{Ceder1998, Meng2009, Islam2014, Stratford2017} 
On the anode side,
fundamental DFT studies have dealt with both the pristine intercalation 
compound LiC$_{6}$ (e.g., Refs.\ \citenum{Kganyago2003} and
\citenum{Toyoura2008})
and with Li adsorption on pristine and defective graphene.
\cite{Khantha2004, Rytkonen2007, Persson2010,
Fan2012, Lee2012a, Zhou2012, Liu2013a, Liu2014b}
For example, in Ref.\ \citenum{Persson2010},
the authors measured the diffusivity of Li in highly ordered pyrolytic 
graphite and compared to theoretical diffusivities,
using the nudged-elastic-band (NEB) method to map out the activation barrier
for an individual atomic jump and feeding these barriers into kinetic
Monte Carlo simulations.

Due to the computational cost and scaling behavior of DFT,
all these simulations are
restricted to relatively small systems,
up to a few hundred atoms at most and short time scales. 
This can be feasible when studying a well-defined 
unit cell (such as in many crystalline oxide cathode materials),
\cite{Islam2014}
but becomes very problematic when attempting to simulate 
disordered or even amorphous
systems (such as carbon nanostructures in anodes),
which require large simulation cells. 
In principle, empirical interatomic potentials, which are
much less computationally demanding,
can be used to describe 
metal atoms in complex environments. \cite{Li2017}
For example, an Li--C parameter set has been developed for the
widely used reactive force field (ReaxFF):\cite{Han2005}
this method has been applied to fracture and failure
mechanisms of carbonaceous electrodes in the presence of Li,
\cite{Yang2013, Huang2013} and, more recently, in a new
implementation, to Li clusters aggregating on graphene surfaces.\cite{Raju2015}
Very recently, ReaxFF was combined with neutron diffraction and pair distribution function
analysis to trace Li atoms in carbonaceous anode
materials in real space. \cite{McNutt2017}

Nonetheless, inherent challenges remain for any empirical
interatomic potential. Examples that are directly relevant to carbon
nanostructures include a poor description of ductile versus
brittle failure in carbon nanotubes (which can be remedied by
appropriate environment-dependent cutoffs)
\cite{Pastewka2008, Pastewka2012}
and the fact that vastly different carbon nanostructures are
obtained from annealing amorphous precursors, depending on
which particular interatomic potential
is chosen. \cite{deTomas2016}

In this work, we introduce an alternative route toward  
atomistic modeling of Li intercalation,
taking the increasingly popular approach of building
accurate yet fast interatomic potentials using
machine-learning (ML) techniques applied to DFT reference data.
We show how Gaussian approximation
potential (GAP) models can be generated by fitting 
to the energy and force {\em differences} that Li atoms
induce in graphitic and amorphous carbon structures.
Rather than focusing on ideal graphite alone, we aim for transferability
and therefore include a large number of disordered and higher-energy
structures.
We analyze the accuracy limits of any difference-based interatomic
potential with a finite cutoff radius,
validate our GAP model against DFT reference data, and discuss
the application to
molecular-dynamics (MD) simulations.

\section{Theory and methods}

\subsection{Machine-learning-based potentials: a brief overview}\label{sec:MLP}

To overcome the limits both of DFT methods and 
empirical interatomic potentials, a popular strategy in 
condensed-phase simulations is to ``machine-learn'' from
DFT reference data and subsequently use this to 
construct a computationally much faster potential.
These ML methods perform a high-dimensional fit to the
DFT potential-energy surface for a limited set of preselected configurations
and then interpolate energies and forces for other structures of interest. 
In contrast to empirical potentials (which are also often
fitted to DFT data), ML models impose no particular functional form and
so can fully flexibly adapt to data; this avoids bias in construction,
but also requires careful fitting and testing to rule out
unphysical behavior.
Over recent years, ML-based 
interatomic potentials using artificial neural networks,
\cite{Behler2007, Artrith2011, Sosso2012, 
Artrith2016, Smith2017, Hajinazar2017, Faraji2017, Kobayashi2017}
Gaussian process regression, 
\cite{Bartok2010, Szlachta2014, aC_GAP}
or other algorithms 
\cite{Botu2015, Seko2015, Li2015, Shapeev2016, Kruglov2017, 
Podryabinkin2017, Huan2017}
have been attracting growing interest.
They were successfully used to 
describe complex atomistic processes, 
such as the crystallization of the phase-change material GeTe
\cite{Sosso2013, Sosso2015a, Gabardi2017}
or various high-pressure phase transitions
in elemental solids. \cite{Behler2008, Khaliullin2011, Eshet2012}
The ability to reach close-to-DFT accuracy at much lower 
computational cost 
makes them particularly promising tools for studying
disordered and amorphous systems.\cite{Sosso2013,aC_GAP}
The current state of the field
has been reviewed, for example, in Refs.\ \citenum{Rupp2015} and
\citenum{Behler2016}.

Notwithstanding their usefulness, ML-based interatomic potentials face challenges as well. Among the most central ones is the need for large reference databases of DFT data that cover many very different scenarios, such as transition states, defects, and surfaces. This challenge becomes particularly pressing with increasing
 chemical complexity: for elemental solids, 
systematic reference databases can be constructed (as discussed,
e.g., in Ref.\ \citenum{Szlachta2014}), but for binary, 
ternary, or quaternary chemical systems
the complexity grows very quickly. Intercalated atoms represent
a special case, which we will discuss in the following.

\subsection{Difference-based fitting for atom intercalation}

Li intercalation, representative of the more general scenario of
guest atoms in host species, involves two elemental components
but not on equal footing. In the present case, a GAP model for the host
(carbon) structure is already available,\cite{aC_GAP} to which we
aim to add Li guest atoms in a second step. This is particularly relevant as the
nonlocality (the expected error in the ML fit) is quite sizeable
for amorphous carbon, on the order of 1 eV \AA{}$^{-1}$
for interatomic forces, imposing a
natural and insurmountable bound on the achievable accuracy of 
finite-range potentials.\cite{aC_GAP}
(Nonetheless, this GAP model enables accurate predictions 
for structural and mechanical
properties of amorphous carbon, as shown in detail in Refs.\ \citenum{aC_GAP}
and \citenum{Laurila2017}.)
By contrast, the force nonlocality that arises from
Li insertion in these structures is much smaller, as 
will be seen in the following.

\begin{figure}[tb]
\centering
\includegraphics[width=\figurewidth]{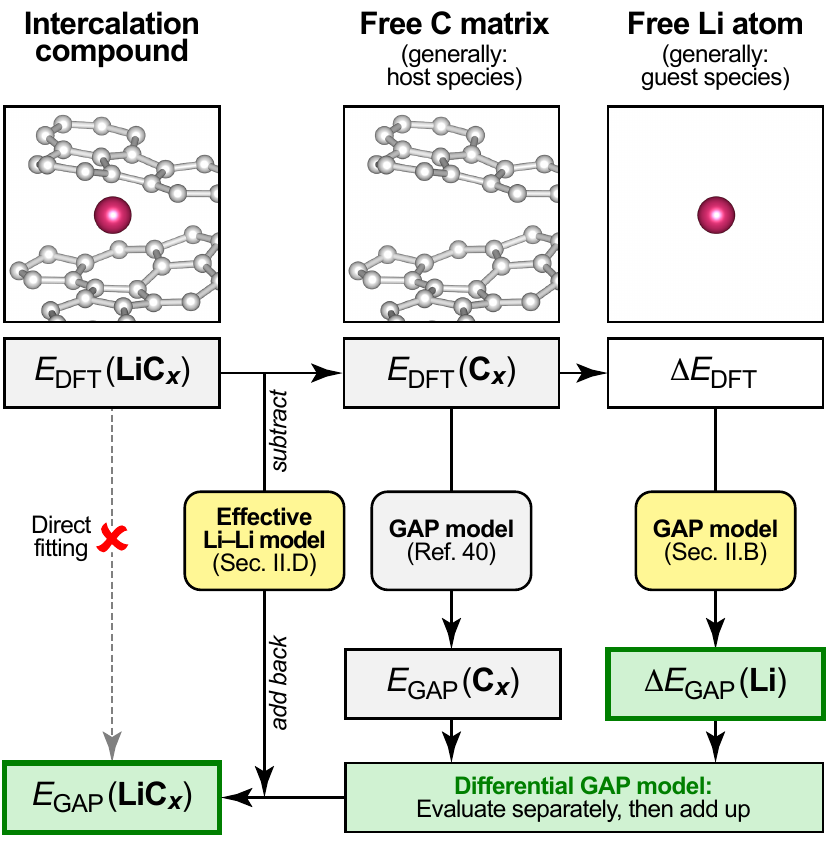}
\caption{\label{fig:overview}
         Overview of the approach employed here. Rather than fitting directly
         to energies (and forces) for LiC$_{x}$ systems, we repeat
         the DFT computations for the same structures without Li, and 
         fit a GAP model to the energy and force differences. The
         latter is combined with a previous GAP model for C--C
         interactions (Ref.\ \citenum{aC_GAP}).
         A baseline model for long-range Li--Li interactions is subtracted
         from the reference dataset prior to fitting, and added back onto
         the output.
         Summing up the terms yields the total GAP energy for the system under
         study, and similar procedures hold for the forces on atoms.}
\end{figure}

Rather than aiming at an explicit description of the full Li--C binary system,
we therefore propose to perform a fit for the energy {\em differences}
arising from inserting an Li atom into a C matrix, as sketched in
Fig.\ \ref{fig:overview}; that is, we seek an ML
representation for the intercalation energy
\begin{align}
\label{eq:DeltaE}
 \Delta E_{\rm DFT} \equiv E_{\rm DFT}({\rm LiC}_{x}) 
                    - E_{\rm DFT}({\rm C}_{x}) 
                    - E_{\rm DFT}({\rm Li})
\end{align}
as a function of the atomic coordinates involved. In addition, the
first derivative of the energy difference yields the force difference,
which makes it feasible to fit a model including forces, and thus increases
the amount of available data.

The energy for the
pure carbon framework, $E_{\rm DFT}({\rm C}_{x})$, is already
accessible through our GAP model for amorphous
carbon. \cite{aC_GAP}
The flexibility of this model has been exemplified very recently by using
it for random structure searching, leading to the identification of 
several hitherto unknown
hypothetical carbon allotropes.\cite{C_AIRSS} 
We tested the quality of the initial, pure carbon GAP specifically for
ten snapshots from MD simulations with guest atoms (see below),
in which we removed the Li atoms and performed
static computations for the remaining, distorted carbon-only
structures: this gave 
a root-mean-square energy error on the order of 0.05 eV/at.~against DFT.
Hence, using this elemental carbon GAP, we can
predict the energy for the pure host framework during a simulation, i.e.,
we have direct access to 
$E_{\rm GAP}({\rm C}_{x})$. Furthermore, the 
energy of an isolated Li atom, $E_{\rm DFT}({\rm Li})$, 
is constant and does not need to be part 
of the ML framework.

In this work, we therefore construct a new GAP model for the energy
{\em difference}, $\Delta E_{\rm DFT}$, using established methods which
will be described below.
Note that while the discussion here focuses on energies for
simplicity, the overall fitting process involves the forces on atoms as
well, and it employs a sparsification procedure to
select only representative atomic environments from the multitude of
configurations in the reference database.\cite{Mahoney2009}
A detailed description of the fitting procedure
is found in Ref.\ \citenum{Bartok2015}.

The ideas outlined here are developed in the GAP framework, but they are
expected to be readily transferable to other implementations of ML-based
potential fitting. Indeed, while this manuscript was in preparation,
we became aware of very recent work by Li et al., \cite{Li2017a} who generated
a neural-network potential for Cu adatoms in amorphous Ta$_{2}$O$_{5}$. In
this case, the host structure is fixed (and its energy obtained directly 
from DFT evaluations), but the idea of fitting to energy {\em differences}
induced by an adatom (there, Cu; here, Li) is very similar. In addition,
Li et al.\ pointed out how such an ML-based potential can be used to 
perform accurate NEB computations, which can subsequently be 
combined with kinetic Monte Carlo
modeling.\cite{Li2017a}

\subsection{Structural descriptors}

How does one ``teach'' chemical structure to an ML algorithm?
This choice of a mathematical prescription for encoding atomic environments,
of so-called ``descriptors'', is indeed crucial for the success of any
ML-based interatomic potential. \cite{Bartok2013}
Many-body descriptors have been successfully used: \cite{Behler2007, Bartok2010}
all neighbors of a given atom are included up to a specified cutoff distance. 
We have recently shown
how non-parametric two- and three-body
(distance and angle) terms can be combined with a
many-body descriptor
in the GAP framework; \cite{aC_GAP} this improves the robustness of the fit,
especially when describing highly
disordered liquid and amorphous structures.
Our GAP model for Li intercalation uses similar ideas but additionally needs to
distinguish Li--C and Li--Li interactions, and utilizes a total of four descriptors:
\begin{itemize}
 \item a two-body term for Li--C interactions;
 \item a two-body term for Li--Li interactions; both use the
       distance between atoms as a simple scalar descriptor coordinate;
 \item a three-body term for the angles that an Li atom forms with two neighboring
       carbon atoms; and finally
 \item a many-body term that includes all C neighbors of a given
       Li atom, up to a specified cutoff radius, $r_{\rm cut}$. 
\end{itemize}

Summing over these four descriptors as indicated by the superscript ``$(d)$'', each expressed through general vectors
${\vec{q}}^{(d)}$, and using a scaling
parameter $\delta^{(d)}$ for the different contributions,
gives the final expression:
\begin{align}
  \label{eq:gap-total-energy}
  \Delta E_{\rm GAP} =& \sum_{d} \left\{
  \delta^{(d)} \sum_{i} \sum_{t} \alpha^{(d)}_{t}
            K^{(d)} \left({\vec{q}}^{(d)}_{i}, {\vec{q}}^{(d)}_{t}
\right) \right\},
\end{align}
where $\alpha^{(d)}_{t}$ are fitting coefficients, and 
$K^{(d)}$ is a similarity measure or {\em kernel} that compares the
$i$-th environment in the trial structure to the $t$-th one in the reference
dataset (from which $N_{t}$ points are drawn).
For two- and three-body interactions, we use a simple squared
exponential kernel,\cite{Bartok2010}
\begin{equation}
 K^{(d)} \left({\vec{q}}^{(d)}_{i}, {\vec{q}}^{(d)}_{t} \right) =
  \exp \left[ -\frac{1}{2} \sum_{\xi} 
  \frac{(q_{\xi, i}^{(d)} - q_{\xi, t}^{(d)})^{2}}{\theta^{2}} \right],
\end{equation}
where the index $\xi$ runs over the individual components of the descriptor
vector, and the parameter $\theta$ controls the selectivity of the kernel.

For many-body interactions, we employ
 the Smooth Overlap of Atomic Positions (SOAP) approach
\cite{Bartok2013} that has been previously used for generating GAP models,
\cite{Szlachta2014, aC_GAP} for restraining refinements of diffraction data,
\cite{Cliffe2017} and for classifying molecular
and condensed-phase structures. \cite{De2016} SOAP expands the
neighbor density around a given atom $a$
into a basis set of orthogonal, atom-centered functions,
\begin{equation}
  \label{eq:soap_rho_expansion}
  \rho_{a}({\vec{r}}) = \sum_{nlm} c^{(a)}_{nlm} \, g_{n}(r) Y_{lm}(\hat{\vec{r}}),
\end{equation}
where $g_{n}(r)$ denote radial basis functions and 
$Y_{lm}(\hat{\vec{r}})$ are spherical harmonics,
up to a specified maximum value of $n$ and $l$
(here, we choose $n_{\max} = l_{\max} = 10$). 
The expansion coefficients $c^{(a)}_{nlm}$ are then
used to form the power spectrum, 
\begin{equation}
 p^{(a)}_{nn'l} =\sqrt{\frac{8 \pi^{2}}{2l+1}} \sum_{m} \left( c^{(a)}_{nlm} \right)^{\ast} c^{(a)}_{n'lm},
\end{equation}
which makes it possible to conveniently evaluate the similarity between
two atomic environments in the form of a dot product:
\begin{equation}
 k({\vec{q}}^{\rm (MB)}_{a}, {\vec{q}}^{\rm (MB)}_{t}) = 
 \sum_{nn'l} p^{(a)}_{nn'l} \, p^{(t)}_{nn'l} = 
 {\vec{q}}^{\rm (MB)}_{a} \cdot {\vec{q}}^{\rm (MB)}_{t}.
\end{equation}

Finally, to better distinguish between different  
environments, we raise this similarity measure to a
small positive power $\zeta$, leading to the final expression
\begin{equation}
 K^{\rm (MB)}({\vec{q}}^{\rm (MB)}_{a}, {\vec{q}}^{\rm (MB)}_{t}) = 
 \left| {\vec{q}}^{\rm (MB)}_{a} \cdot {\vec{q}}^{\rm (MB)}_{t} \right|^{\zeta}
\end{equation}
(here, we choose $\zeta = 4$).
The remaining
parameters used for the GAP model are provided in Table \ref{tab:GAP-params}.

\subsection{Effective Li--Li potential}\label{sec:Veff}

Up to this point, we have discussed the fitting to energy and force
differences (Eq.~2; Sec.~II.B), and we tried to combine this with previous GAP
fitting strategies that were successfully used for our elemental carbon
model (Sec.~II.C). \cite{aC_GAP} However, direct application of these methods did not
lead to a satisfactory description of Li--Li dynamics (see below), and we
found an additional methodological step to be necessary.

Despite getting satisfactory fits to $\Delta E_{\rm DFT}$, the source of the problem was traced back to the fact that in terms of absolute value, the largest contribution to $\Delta E_{\rm DFT}$ is coming from individual Li insertion energies (typically $>\!1$~eV), but the dynamics of Li atoms is governed to a significant extent by Li--Li interactions, which are comparatively much weaker (typically $\sim 0.1$~eV). This weaker interaction is difficult to tease out from the data. Therefore we introduce an effective Li-Li interaction term, which we calculate with DFT explicitly, and fit directly with a pair potential (we use a two-body GAP term for this). This effective potential is subtracted from the DFT data prior to fitting the rest of the model, and then added back to obtain the full ML model (Fig.\ \ref{fig:overview}). Thus, the end result is an accurate fit to  $\Delta E_{\rm DFT}$, but which also has an explicit term that is designed to capture effective Li--Li pair interactions, including the long-range behavior, as well as possible.

Let us consider a carbon framework in which two Li atoms (``A'' and ``B'')
are intercalated.
The total energy of this system, $E_{\rm AB}$, is accessible via DFT, 
and we  decompose it into the energy of the Li-free
system, $E_{\rm free}$, and a number of additional terms induced by the
guest atoms:
\begin{align}
 E_{\rm AB} = E_{\rm free} + \delta E_{\rm A} + \delta E_{\rm B} + V_{\rm eff}^{\rm Li-Li},
\end{align}
where $\delta E_{\rm A}$ and $\delta E_{\rm B}$ correspond to changes in energy of the system due to the presence of either Li atom on its own, viz.
\begin{align}
\nonumber \delta E_{A} \equiv E_{\rm A} - E_{\rm free},\\
\nonumber \delta E_{B} \equiv E_{\rm B} - E_{\rm free},
\end{align}
and the final term is the effective Li--Li interaction potential, which
is precisely what we are looking for. Rearranging gives
\begin{align}
 \nonumber
 V_{\rm eff}^{\rm Li-Li} =& E_{\rm AB} - E_{\rm free} - \delta E_{\rm A} - \delta E_{\rm B}\\
 \nonumber
 =& E_{\rm AB} - E_{\rm free} - \left[E_{\rm A} - E_{\rm free}\right] 
     - \left[E_{\rm B} - E_{\rm free}\right] \\
 =& E_{\rm AB} - E_{\rm A} - E_{\rm B} + E_{\rm free}.
 \label{eq:Veff}
\end{align}

\begin{figure}
\centering
\includegraphics[width=\narrowfigurewidth]{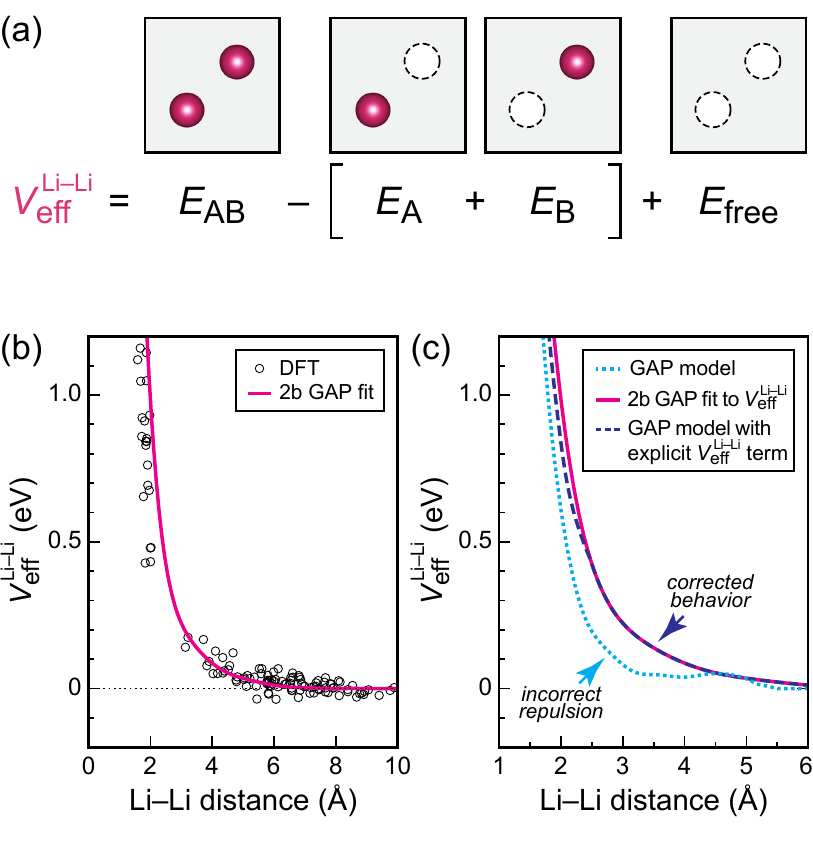}
\caption{\label{fig:corePotential}
         An effective, machine-learned potential for Li--Li interactions.
         (a) Schematic illustration of how the effective potential can be
         obtained from sets of DFT computations with atoms A and B either
         present or absent (Eq.\ \ref{eq:Veff}).
         (b) Results of this evaluation for a range of structures with
         different A--B distances, showing
         original DFT data (circles) and a two-body (2b) GAP model 
         that has been fitted to
         these data (magenta line).
         (c) Potential-energy scan for two Li atoms in vacuum, as a function
         of their interatomic distance. The dotted, light blue line shows the
         result of a direct GAP fit (not including the effective
         potential)---this
         underestimates the repulsion at around
         3 \AA{}, and therefore led to spurious Li--Li correlations in
         preliminary simulations
         ({\em cf.} Fig.~\ref{fig:MD216} below).
         Including $V_{\rm eff}^{\rm Li-Li}$ in
         the full model (dashed, navy line) alleviates this problem. 
}
\end{figure}

In other words, an effective
Li--Li potential can be extracted from sets of
DFT computations which have pairs of atoms present (AB), one or the other
removed (A/B), and finally both removed (``free''), all for the same carbon framework
(Fig.\ \ref{fig:corePotential}a). Similar expressions can be derived for
the forces on atoms.

We performed sets of such DFT computations, taking care to sample various
Li--Li distances, both the strong repulsion at around 2 \AA{} of separation and the longer-range behavior beyond 3 \AA{}. We then fitted a 2-body GAP model to the combined data
(Fig.\ \ref{fig:corePotential}b). 
Note that these two datasets were performed using CASTEP and VASP,
respectively. This does not lead to an inconsistency since only energy
differences are used here, and  
the absolute energy data do not enter this fit.

We stress the simplified nature of this
potential: it is fitted using a two-body descriptor only, that is, it depends
only on the distance between two Li atoms. Its cutoff is 9 \AA{}, significantly
longer than typically used values for short-range GAPs.

This effective potential is subtracted from the input data that
enter the differential fit (Eq.\  \ref{eq:GAP}), and subsequently it is 
added back to give the final, corrected result. 
We illustrate the need for
this procedure in Fig.\ \ref{fig:corePotential}c. There, we have performed a
simple diagnostic test by computing
the interaction energy of two free Li atoms. An initial version of the
GAP model (not including $V_{\rm eff}^{\rm Li-Li}$; dotted, light blue line)
significantly
underestimates the repulsion at 2--3 \AA{} separation. In consequence, running
GAP-driven MD simulations with this preliminary
model led to an incorrect behavior in the Li--Li
radial distribution function---that is, to an unphysical Li--Li attraction at
distances at up to 4 \AA{}. By
including $V_{\rm eff}^{\rm Li-Li}$ in the full GAP model,
the physical behavior is correctly recovered (dashed, navy line).

The final expression for the energy in our combined GAP model hence reads
\begin{align}
 \nonumber
 E_{\rm GAP}({\rm LiC}_{x}) =&  \,
                            E_{\rm GAP}({\rm C}_{x}) + E_{\rm DFT}({\rm Li})\\
                            & + V_{\rm eff}^{\rm Li-Li}  + \Delta E_{\rm GAP};
                             \label{eq:GAP}
\end{align}
that is, we approximate the intercalation energy as
\begin{align}
 \Delta E_{\rm DFT} \approx V_{\rm eff}^{\rm Li-Li} + \Delta E_{\rm GAP}.
\end{align}

We note the analogy of the above to how molecular solids and
liquids are treated using the molecular many-body expansion.
\cite{Bartok2013a}
Beyond this particular system,
we believe that such approaches will be of more general interest for those
ubiquitous scenarios where relatively weak interactions need to be
treated in ML-based materials simulations.

\begin{table}[tb]
\centering
\caption{Key parameters for the GAP models created in this work
         (as outlined in Fig.\ \ref{fig:overview}):
         we first fit an effective Li--Li two-body potential (Sec.\ \ref{sec:Veff};
         parameters given in italics), and then fit a combined model for
         $\Delta E$ to data from which $V_{\rm eff}^{\rm Li-Li}$ has been
         subtracted.
         The notation follows the definitions in the text.
         $r_{\Delta}$ is a transition
         width for SOAP;\cite{Bartok2013}
         $\xi$ denotes the dimensionality of the descriptor
         (i.e., its number of components),
         and $\theta$ and $\sigma_{\rm at}$ control the smoothness 
         of the respective kernels.}
\label{tab:GAP-params}
\begin{tabular}{lccccc}
\hline
\hline
 &   & \multicolumn{4}{c}{$\Delta E$ model} \\
 \cline{3-6}
 & $\,$  $\,$ & $\,$ 2-body $\,$ & $\,$ 2-body $\,$ & $\,$ 3-body $\,$ & 
                       SOAP \\
 & $V_{\rm eff}^{\rm Li-Li}$ & Li--Li & Li--C & C--Li--C & Li--C \\
\hline
$r_{\rm cut}$ (\AA{})  & {\em 9.0} & 2.5 & 5.5 & 3.5 & 4.5  \\
$r_{\Delta}$  (\AA{})  & &     &     &     & 0.5  \\
$\xi$                  & {\em 1} & 1   & 1   & 3   & 606  \\
\hline
$\delta^{(d)}$         & {\em 1} & 10  & 1 & 0.1 & 0.1   \\
\hline
$\theta$               & {\em 2.2} & 0.8 & 0.8 & 1.2 &      \\
$\sigma_{\rm at}$ (\AA{}) & &  &     &       & 0.5  \\
\hline
$N_{t}$ (total)        & {\em 32} & 12 & 20 & 200 & 3500 \\
\hline
\hline
\end{tabular}
\end{table}

\section{Computational details}

\subsection{Training database and model fitting}

Initial training data were generated by randomly placing Li atoms in
slightly (randomly) distorted graphite
(24 atoms/cell), graphene (24 atoms/cell), and amorphous carbon
(64 atoms/cell) structures (maximum Li concentration 10 atomic\ \%). 
The latter were generated by quenching from the melt at different
densities following Ref.\ \citenum{aC_GAP} and using the GAP model
introduced there.
Single-point DFT computations were performed for all these structures:
as we need the energy and force differences for fitting, reference computations
for the Li-free structures were also required. To save computational time, one 
single Li-free structure can be used to generate several structures with
one or several Li atoms present.

In most cases, we enforced a minimum Li--C distance of 1.5 \AA{} in the
training data (``hard-sphere constraint'');
to accurately describe high-energy structures, we
included a small amount of data where the minimum distance was lower,
down to 0.70 \AA{}. 
We attempted to sample configuration space widely;
however, structures in which unphysically high energies ($> 20$ eV/Li)
or forces ($> 80$ eV \AA{}$^{-1}$) occurred were excluded (after testing
several limits).
The final training set contains
561 graphite, 192 graphene, and 1664 amorphous configurations. 

Based on these training data, the GAP was fitted in an iterative fashion. We
started by a model that only contains a two-body descriptor for Li--C interactions,
and chose the parameters so as to strike a compromise between accuracy and
simplicity (that is, achieving reasonable values for
$r_{\rm cut}$ and $N_{t}$).
Once these were optimized, the next descriptor
was added (Table \ref{tab:GAP-params}); this process was repeated until the 
introduction of new descriptors led to no further improvements.
The potential parameter files and DFT reference database are available as described in the Data Access Statement at the end of this paper; the GAP prediction and training codes are available at http://www.libatoms.org.

\subsection{DFT computations}

Reference DFT computations, both for the fitting of the potential and its
validation, were carried out in the local-density
approximation (LDA), which had been used for the initial carbon potential due to its conceptual simplicity and its satisfactory description of the graphite interlayer distance.\cite{aC_GAP} In future work, it will likely be beneficial to explore the effect of advanced dispersion-correction methods such as many-body dispersion corrections,\cite{Gobre2013} but this does not affect the questions and concepts under study in this work. It is further known that the description of Li intercalation
in graphite can be further improved by higher-level DFT and by 
Quantum Monte Carlo (QMC) methods. \cite{Ganesh2014}
These are not the target here, however, due to their much higher computational
cost, and since the fitting
procedure is independent of the underlying DFT methodology.
With more computational power available,
it should be possible to fit to these quantum-mechanical methods in the
future.

Single-point energy and force
computations were carried out using CASTEP,\cite{CASTEP}
following protocols in our previous work. \cite{aC_GAP}
Reciprocal space was sampled on meshes with a maximum spacing of 0.03 \AA{}$^{-1}$.
The halting criterion for SCF iterations was $\Delta E < 10^{-8}$ eV.
Pseudopotentials were generated on-the-fly, 
with a plane-wave energy cutoff of 650 eV,
and a correction for finite-basis errors was employed. \cite{FBSC}

To obtain reference data for dynamical properties of interest, viz. radial
and bond-angle distributions as well as vibrational densities of states, additional
DFT-based MD simulations were performed using the Vienna
Ab initio Simulation Package (VASP),\cite{Kresse1993, Kresse1996a, Kresse1999}
the projector augmented-wave method, \cite{Blochl1994} and the LDA.
In these simulations, the Brillouin zone was sampled at the $\Gamma$ point, which is standard
practice for obtaining long MD trajectories.
The plane-wave energy cutoff was 500 eV.
The temperature was set to 1,000 K and
controlled by a Nos\'e{}--Hoover thermostat. 
MD simulations were performed for 75 ps with a time step of
1 fs; the last 50 ps of the trajectories were sampled for analysis.
The same or similar settings were chosen for GAP-driven MD 
simulations wherever possible. 

\section{Results and discussion}

\subsection{Locality and target accuracy}\label{sec:locality}

\begin{figure}
\centering
\includegraphics[width=\figurewidth]{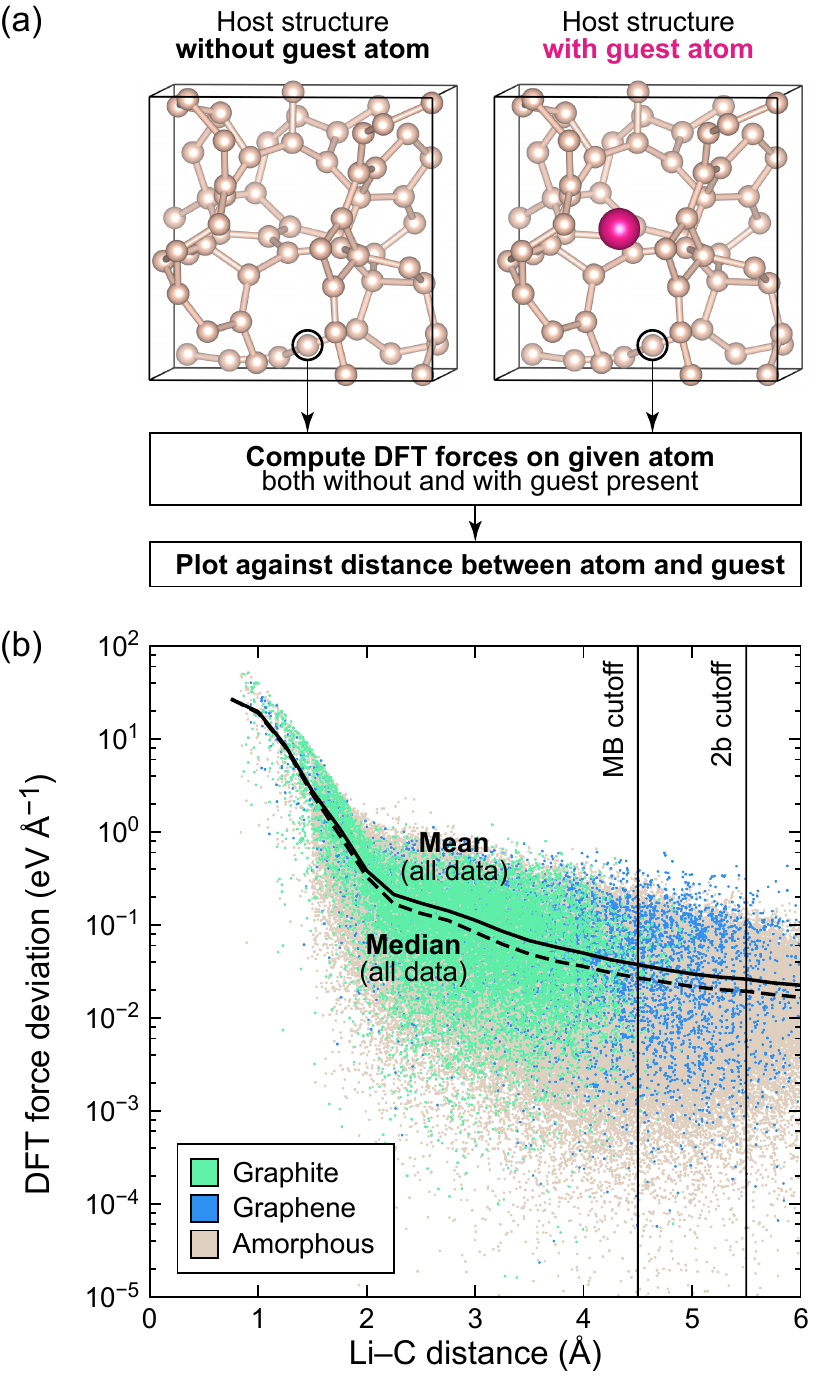}
\caption{\label{fig:locality}
         Locality tests for guest atoms in host structures: quantifying
         the maximum possible accuracy for a finite-range interatomic
         potential. 
         (a) Sketch of the procedure using an exemplary amorphous carbon
         structure.
         (b) Results for a large database of graphite (green), graphene
         (blue), and amorphous carbon (gray) structures, with and without
         a single Li atom inserted in each. The mean
         (solid line) and median (dashed line) are collected with a
         binning interval of 0.25 \AA{}. 
}
\end{figure}

Our potential is fitted with the assumption of locality: interactions
beyond a given cutoff radius are not part of the model, and therefore
a natural bound is placed on how accurate any finite-range
potential can be. We have previously demonstrated
how locality can be assessed for crystalline and amorphous compounds:
by defining a fixed sphere around an atom in a structure, perturbing all
atoms outside this sphere, and measuring the forces on the central atom
as a function of sphere size.
\cite{Bartok2010, aC_GAP} Here,
we are interested in the force differences due to Li insertion,
and therefore have an easier way of quantifying locality: we inspect the force
components on each carbon atom in a structure with and without a single Li atom intercalated
(Fig.\ \ref{fig:locality}a), and plot the difference (for each of the three
Cartesian force components individually) as a function of how far
this particular atom is away from the intercalated Li.

To make the interpretation of the data easier, we collect them with
a binning interval of 0.25 \AA{} and calculate the arithmetic mean and
median values for each bin (solid and dashed lines in Fig.\ \ref{fig:locality}b);
as their behavior is qualitatively similar, we focus on the mean in the
following. 
The trends with Li--C distance reveal two distinct regimes:
up to 2 \AA{}, the
mean deviation (which we take to indicate the expected force error of the
potential) drops steeply but remains very high, as this is the region
of strong Li--C interactions, not of typical cutoff radii. 
From 2 \AA{} onwards, the mean still 
declines, indicating a remaining degree 
of nonlocality in the system,
which would require a potential with a cutoff of $> 6$ \AA{}. However,
this value must be chosen as a compromise: too
large cutoffs will drastically increase the amount of
required DFT reference data and also the complexity of the GAP 
(making it more computationally expensive in runtime). 
The cutoffs we use are indicated by a vertical line: note that
the two-body (pairwise) descriptor extends wider than the many-body (SOAP) 
descriptor, and we only expect the potential to reach the accuracy
relating to the latter.

\begin{figure}
\centering
\includegraphics[width=\figurewidth]{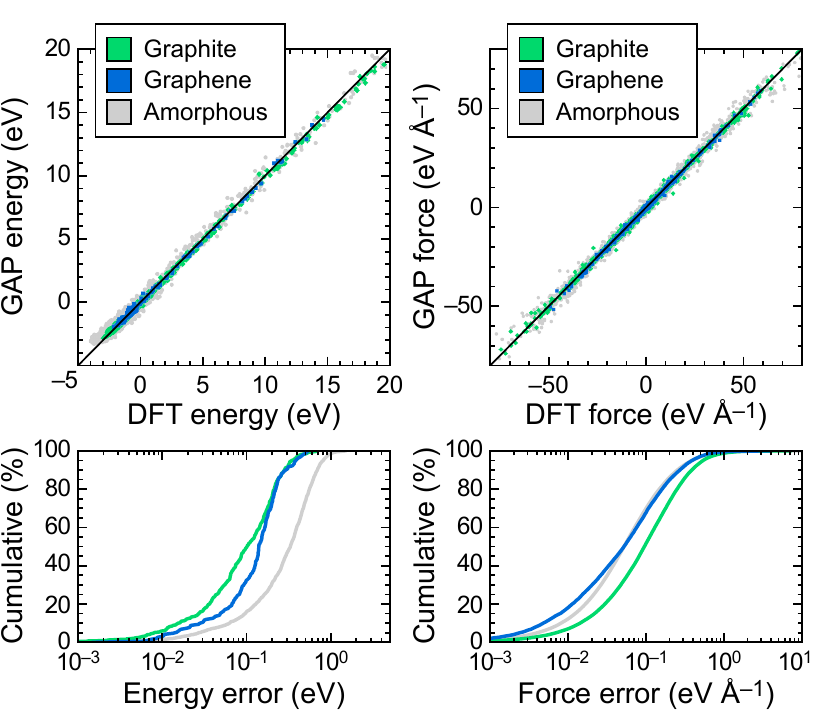}
\caption{\label{fig:scatter}
         {\em Top:} Numerical validation of the GAP, using energies ({\em left})
         and the Cartesian components of forces on atoms ({\em right}),
         as compared to DFT reference data.
         {\em Bottom:} Cumulative distributions for Li insertion 
         energy errors per atom
         and force component errors, respectively. 
         A point on these
         lines means that $y$ percent of the data exhibit an error of $x$
         or less: the further left, the lower the error; the higher up, 
         the higher the confidence.
}
\end{figure}

This locality analysis, performed for the three different
types of configurations individually (Fig.\ \ref{fig:locality}b),
reveals similar trends but a different
spatial extent: this simply mirrors the size of the supercells used,
which are smallest for graphite (green) and largest
for amorphous carbon (gray). Inspecting the mean as in Fig.\ \ref{fig:locality}b,
but now for each set and at around 4.0 \AA{},
we find deviations of
0.10, 0.08, and 0.05 eV \AA{}$^{-1}$ 
for graphene,  graphite, and amorphous configurations,
respectively. Hence, intercalation in graphitic structures
shows slightly higher nonlocality; this is qualitatively 
consistent with previous findings for different forms of elemental
carbon. \cite{aC_GAP}

\subsection{Numerical errors}

The most straightforward test for the quality of the GAP 
(or any ML potential)
is computing
energies and forces for the DFT database and comparing them
point-by-point 
to the reference values.
In the present work, we are studying an intercalation system, and
since the host framework can be sufficiently described by the initial 
pure-carbon potential (Sec.~II.B), we will here focus on the numerical
errors for Li intercalation energies. We will show throughout this paper
that, even in the presence of a notable residual numerical error, our
potential can predict physical properties correctly.

The resulting energy (per atom)
and force scatterplots are presented in
Fig.\ \ref{fig:scatter}. 
Although the overall correlation appears to be satisfactory, the
numerical errors for the Li intercalation energy are notable:
the root-mean-square (mean absolute) energy errors are 0.37 (0.29) eV/at.,
respectively.
We re-iterate that the DFT reference database includes
a large number of amorphous configurations on purpose, as well as 
small Li--C distances (below 1 \AA{} in a few cases; Fig. 3b). Indeed,
for the amorphous subset of DFT data, the errors are highest (RMSE of
0.43 eV/at.), whereas for the graphite and graphene subsets they are 
lower (0.17 and 0.19 eV/at., respectively).
For the present proof-of-concept study, it was our target to sample
configuration space as broadly as
possible, and to show that this set of training data suffices to
recover the dynamics of Li atoms in graphitic-like frameworks.
In future work, it may be interesting to fit to larger databases sampled
from GAP-driven MD trajectories, and thus to sample a more
constrained region of configuration space, in turn achieving 
higher numerical accuracy.

Importantly, despite the notable residual energy error, 
the present version of the GAP performs very well in reproducing
dynamical properties in MD trajectories, as will be shown in Sec.~IV.C below.
We believe that this is partly due to a satisfactory reconstruction of the
interatomic forces. The root-mean-square and mean absolute errors
are 0.25 and 0.12 eV \AA{}$^{-1}$ (Fig.~4b), respectively, and are significantly better
than those for our initial carbon GAP (on the order of 1 eV \AA{}$^{-1}$; Ref.\ \citenum{aC_GAP}). This can be compared to a mean absolute error of 0.06 eV \AA{}$^{-1}$ for a state-of-the-art ML model tested on distorted crystal structures of Al,\cite{Botu2015} 
or to an RMSE of 0.46 eV \AA{}$^{-1}$ for a highly succcessful neural-network
potential for amorphous GeTe.\cite{Sosso2012}
Naturally, the more distorted and diverse the local atomic environments, the larger the overall error will become. We stress that even
with a residual force error, the previous potential had afforded very accurate
predictions of structural, mechanical, and surface properties.\cite{aC_GAP} Indeed, looking at
numerical errors alone appears to be not enough when benchmarking effective potentials
for amorphous
materials.

\begin{figure}
\centering
\includegraphics[width=\narrowfigurewidth]{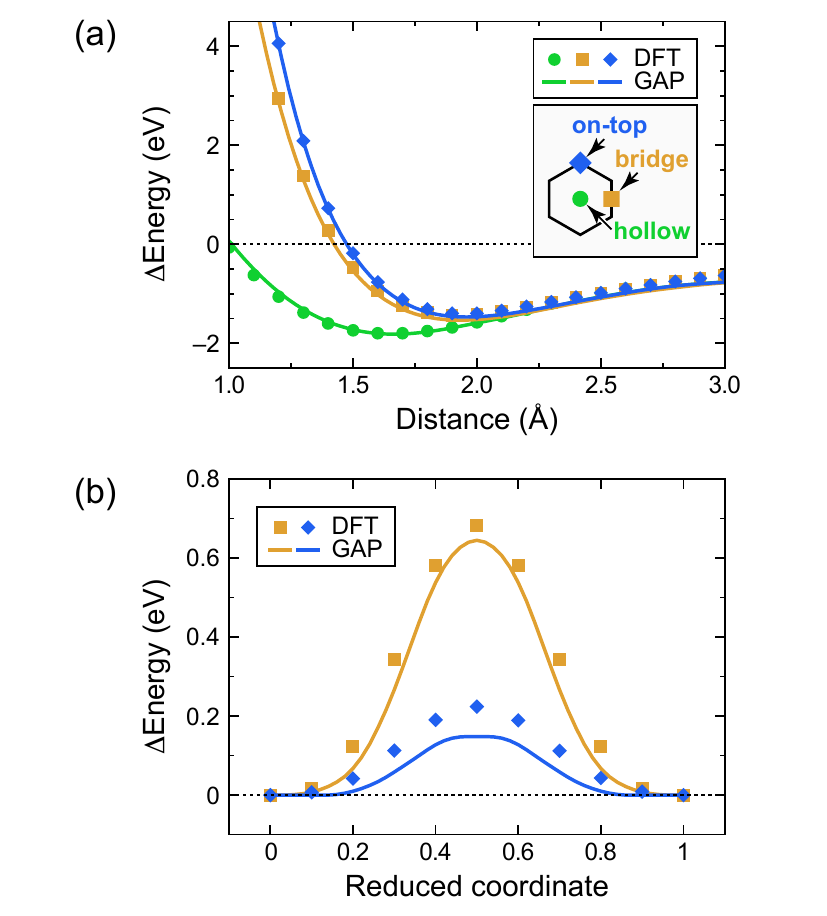}
\caption{\label{fig:graphene}
         Tests of the GAP (line) versus DFT (symbols) for two
         basic atomistic processes: the adsorption of an Li atom
         on high-symmetry sites of a graphene sheet (a), and the diffusion
         of an Li atom between two potential-minimum sites in graphite
         across two high-symmetry pathways (b).
}
\end{figure}

As less comprehensive but more demonstrative tests, 
we next computed characteristic energy profiles for the most fundamental
atomic-scale mechanisms in the Li--graphite system.
We traced the energy profiles of an 
Li atom that is adsorbed on different high-symmetry sites of a graphene
sheet, and of an Li atom that diffuses through pristine graphite,
performing DFT computations for 
reference that are not included in the fit.
All three high-symmetry adsorption sites are correctly captured by the GAP
(Fig.\ \ref{fig:graphene}a), including the
clear preference for the hollow site, and the behavior both at short and long
distances is correctly reproduced. 
The two diffusion pathways in graphite are also qualitatively
correctly described (Fig.\ \ref{fig:graphene}b), albeit
a deviation from the DFT data is visible; as the overall energy differences
involved are smaller, the relative error is slightly more pronounced.
Still, this result is satisfactory as the GAP has to reconstruct the
pathway based on the training data, which do not include the
precise pathway itself.
It is also noted that our numerical tests are highly simplified, by
assuming a perfectly ordered graphite structure; in contrast, experimentally
determined diffusion activation energies
of Li in different carbonaceous materials span a wide range (see, e.g.,
Ref.\ \citenum{Liu1996}).

\subsection{Molecular-dynamics simulations}\label{sec:MD}

The most important question, however---and one that is very difficult to
describe with DFT---, is the description of diffusion through molecular-dynamics
(MD) simulations. 
Recall that for highly ordered systems, diffusivities have been extracted from
NEB simulations of individual jumps, \cite{Persson2010, Islam2014}
using the Arrhenius equation, but this
is not easily 
possible in disordered and amorphous structures as there is a plethora of
different pathways to be considered, each with their own associated barrier. 
A seminal study described DFT-driven MD simulations of Li intercalation in 
carbon nanotubes, but also pointed out the limitations of the 
method.\cite{Meunier2002}
This underlines why new and flexible interatomic
potentials are needed for such applications. 

\begin{figure*}
\centering
\includegraphics[width=14cm]{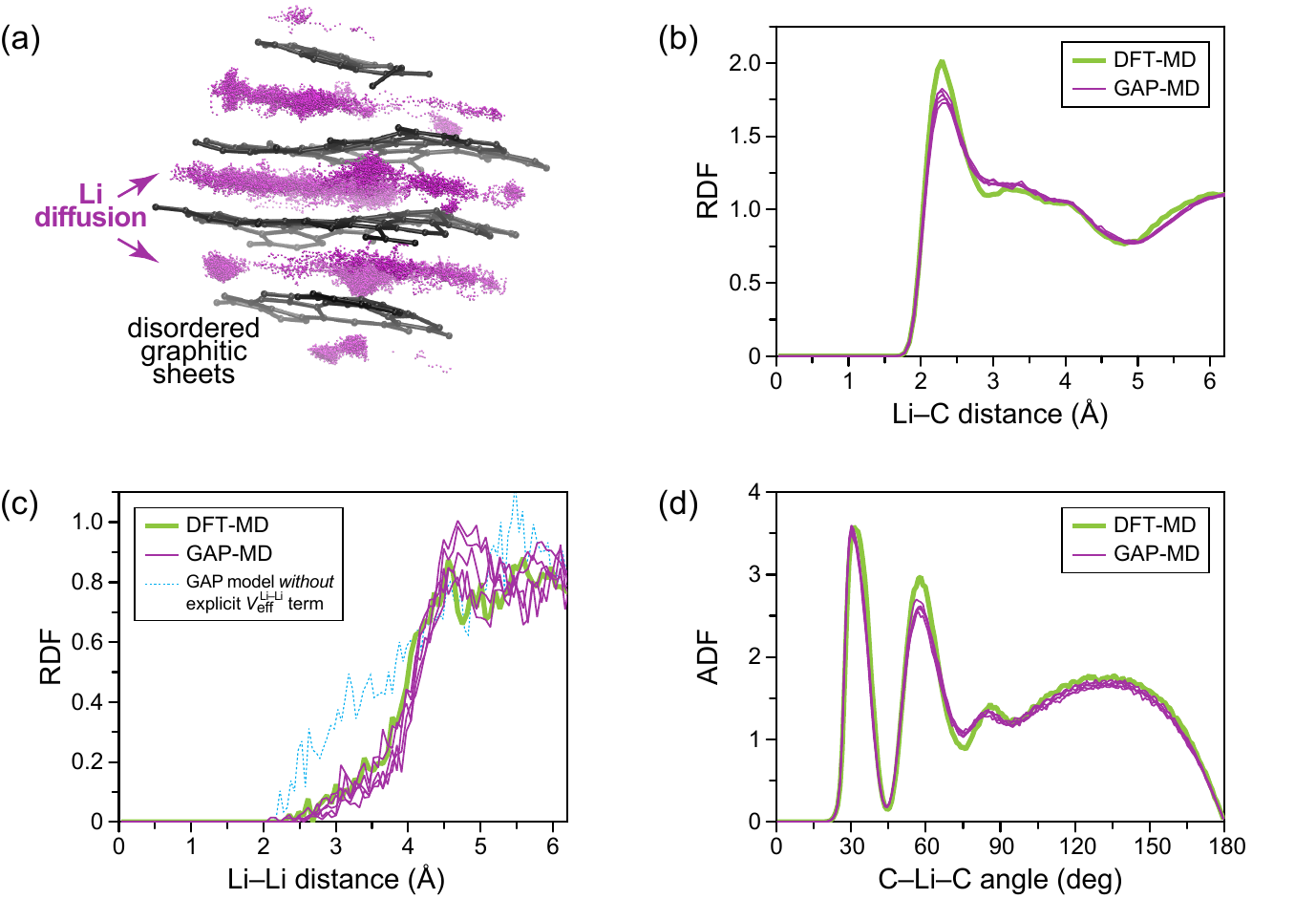}
\caption{\label{fig:MD216}
         Molecular-dynamics simulations of Li diffusion in a graphite-like
         framework at 1,000 K. Results of GAP-MD have been structurally
         benchmarked against DFT-MD data.
         (a) Exemplary GAP-MD trajectory, visualized by plotting 
         coordinates of Li atoms at equally spaced time intervals as 
         purple dots, whereas the carbon framework is shown at a single
         time step to ease visibility. 
         (b) Radial distribution function (RDF) analysis for Li--C contacts
         in this DFT-MD trajectory (green)
         and for five separate GAP-MD trajectories
         computed in the same structure (purple).
         (c) Same for the Li--Li RDF. One exemplary dataset is shown for a
         GAP-MD simulation using a potential {\em without} an effective
         Li--Li potential included (Sec.~\ref{sec:Veff}), and this clearly
         evidences overestimated Li--Li interactions at distances up
         to 4 \AA{} (dashed light blue line).
         (d) Same for the angular distribution function (ADF), determined
         for all C--Li--C angles with a maximum bond length of 3.0 \AA{}.
}
\end{figure*}

To directly assess the performance of our GAP,
we performed MD simulations of an ensemble of four Li atoms in a 
disordered graphite-like structure (Fig.\ \ref{fig:MD216}a); 
the latter consists of predominantly sp$^{2}$-bonded
sheets, reminiscent of graphite but with several defects
(five- and seven-membered rings, as well as a
covalently bonded link between two sheets). 
\footnote{This structure was generated by annealing a disordered 
  amorphous carbon structure using GAP, inducing graphitization 
  following the general ideas in R.\ C.\ Powles, N.\ A.\ Marks, 
  and D.\ W.\ M.\ Lau, Phys.\ Rev.\ B {\bf 79},  075430 (2009). 
  A more detailed account of this will be published elsewhere.} 
This provides us with a suitable test system which allows us to 
probe the interaction of Li atoms with a diverse range of environments;
the simulation cell is small enough to be amenable to DFT computations,
and so we can generate benchmark results for properties of interest.
While we have only been able to obtain a single DFT-MD
trajectory, we performed several parallel GAP-MD runs due to the much lower
computational cost; 
each of these started from the same structure but with different initial
velocities. 
  
To probe the interaction of Li atoms both with the carbon framework
and with one another,
we inspect the radial distribution 
function (RDF) and angular distribution function (ADF) curves
(Fig. \ref{fig:MD216}b--d). The agreement for both 
is highly satisfactory, especially given a certain inherent
scatter in the GAP data (which is not a consequence of the method but
of the system size) and the fact that our model has only 
been trained on small idealized graphite and graphene configurations as 
well as on fully amorphous structures.

The Li--C RDF (Fig. \ref{fig:MD216}b) is the most direct structural
``fingerprint'' of Li intercalation: it shows a maximum at around
2.3 \AA{} both in DFT- and GAP-driven MD.
Small differences remain, in that
the GAP-derived peak is slightly less pronounced; this
is concomitant with a lowering of the average coordination number, obtained
by integrating over the first RDF peak up to 2.6 \AA{}, from 7.3 (DFT)
to 6.9 (GAP). 
The Li--Li RDF (Fig.\ \ref{fig:MD216}c), likewise, is an important
quality criterion: it is where we observed the need for including an
effective Li--Li potential in the final GAP model (Sec.\ \ref{sec:Veff}),
and the latter correctly reproduces the very low likelihood of finding
two Li atoms within 4 \AA{} of one another.
By contrast, the erroneous behavior already discussed above is clearly
seen at the hand of one exemplary GAP-MD trajectory driven by a model
without this effective potential (dashed, light blue line).
In addition to these RDF analyses, the ADF in Fig.\ \ref{fig:MD216}d is a more
complex structural indicator, and is also 
very satisfactorily reproduced by GAP-MD.

We assume that the remaining small differences, in part, may be due to
likewise small
differences in the underlying DFT methods: different implementations and
pseudopotentials are used for generating the GAP reference data and the
DFT-MD trajectory.
Still, the GAP reproduces all general
structural features.

\begin{figure}
\centering
\includegraphics[width=\narrowfigurewidth]{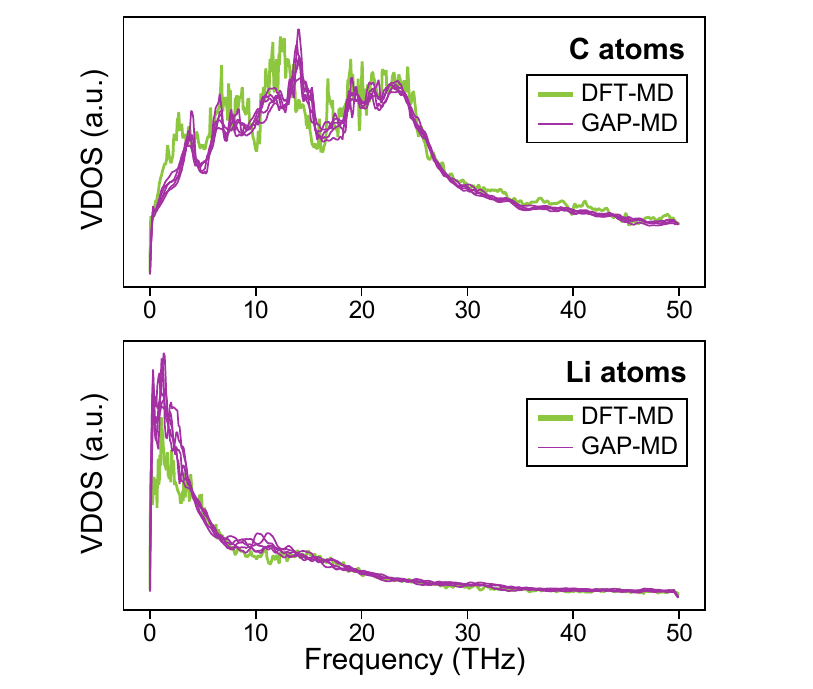}
\caption{\label{fig:VDOS}
         Vibrational densities of states (VDOS), plotted for individual
         MD trajectories as in Fig.\ \ref{fig:MD216}b--c, and separately
         for the host structure ({\em top}) and guest atoms ({\em bottom}). 
}
\end{figure}

As a final means of validation, we extract from the trajectories the
vibrational densities of states (VDOS), using the velocity--velocity
autocorrelation function. This provides information about the atomic
motion in the simulations and a link between local structure
and diffusion dynamics. We inspect the VDOS individually, 
both for the host framework and the Li atoms, in Fig.\ \ref{fig:VDOS}.
There is good agreement between DFT and GAP data, 
and the general features of the VDOS are well reproduced
by our model.
This is particularly so in the 
higher-frequency range ($> 15$ THz), which relates mostly to interatomic
interactions (such as bond-stretching vibrations). At lower 
frequencies, we observe small discrepancies, while the general
trends are preserved. This frequency range is commonly associated
with the diffusion process, and so the above can be understood 
considering the short run-time of the calculation and especially
the small size of the systems (again, both are due to the inherent
computational and scaling limitations of the DFT benchmark, and do
not change the principal validity of our tests).

\section{Conclusions}

Machine-learning-based interatomic potentials for guest atoms in host
structures can be created by fitting to the energy and force differences
which they induce.
We exemplified this for Li intercalation in graphitic and disordered
carbon structures, using the GAP framework to construct an interatomic
potential model.
Notwithstanding notable remaining numerical energy errors, 
reaching up to $\approx 0.4$ eV/atom for Li insertion, the potential
shows satisfactory force accuracy and good transferability, and it
can correctly describe the structural and vibrational properties of
Li diffusion in a carbonaceous framework during high-temperature
molecular-dynamics simulations.
The approach is expected to be more general and
can likely be applied to other classes of ML-based
potentials. 
In the long run, this promises to establish
difference-based ML potentials as a useful simulation method for
electrochemistry and other technologically relevant applications.

\section*{Supplementary Material}
Potential parameter files are provided as Supplementary Material.

\begin{acknowledgments}
S.F.\ gratefully acknowledges support from a French Government
Scholarship and the Foundation of \'E{}cole des Ponts ParisTech,
as well as from Prof.\ Wantanabe at the Department of Materials
Engineering, University of Tokyo, who kindly provided access to
computational facilities in his laboratory.
V.L.D.\ gratefully acknowledges a Feodor Lynen fellowship
from the Alexander von Humboldt Foundation, a Leverhulme Early
Career Fellowship, and support
from the Isaac Newton Trust. 
Computational support was provided by the UK national 
high-performance computing service, ARCHER, for which access 
was obtained via the UKCP consortium and funded by EPSRC grant 
EP/K014560/1.
{\em Data access statement:} Data supporting this
publication will be made available through an online repository
after acceptance.
\end{acknowledgments}


\begin{thebibliography}{61}%
\makeatletter
\providecommand \@ifxundefined [1]{%
 \@ifx{#1\undefined}
}%
\providecommand \@ifnum [1]{%
 \ifnum #1\expandafter \@firstoftwo
 \else \expandafter \@secondoftwo
 \fi
}%
\providecommand \@ifx [1]{%
 \ifx #1\expandafter \@firstoftwo
 \else \expandafter \@secondoftwo
 \fi
}%
\providecommand \natexlab [1]{#1}%
\providecommand \enquote  [1]{``#1''}%
\providecommand \bibnamefont  [1]{#1}%
\providecommand \bibfnamefont [1]{#1}%
\providecommand \citenamefont [1]{#1}%
\providecommand \href@noop [0]{\@secondoftwo}%
\providecommand \href [0]{\begingroup \@sanitize@url \@href}%
\providecommand \@href[1]{\@@startlink{#1}\@@href}%
\providecommand \@@href[1]{\endgroup#1\@@endlink}%
\providecommand \@sanitize@url [0]{\catcode `\\12\catcode `\$12\catcode
  `\&12\catcode `\#12\catcode `\^12\catcode `\_12\catcode `\%12\relax}%
\providecommand \@@startlink[1]{}%
\providecommand \@@endlink[0]{}%
\providecommand \url  [0]{\begingroup\@sanitize@url \@url }%
\providecommand \@url [1]{\endgroup\@href {#1}{\urlprefix }}%
\providecommand \urlprefix  [0]{URL }%
\providecommand \Eprint [0]{\href }%
\providecommand \doibase [0]{http://dx.doi.org/}%
\providecommand \selectlanguage [0]{\@gobble}%
\providecommand \bibinfo  [0]{\@secondoftwo}%
\providecommand \bibfield  [0]{\@secondoftwo}%
\providecommand \translation [1]{[#1]}%
\providecommand \BibitemOpen [0]{}%
\providecommand \bibitemStop [0]{}%
\providecommand \bibitemNoStop [0]{.\EOS\space}%
\providecommand \EOS [0]{\spacefactor3000\relax}%
\providecommand \BibitemShut  [1]{\csname bibitem#1\endcsname}%
\let\auto@bib@innerbib\@empty
\bibitem [{\citenamefont {Wu}\ \emph {et~al.}(2003)\citenamefont {Wu},
  \citenamefont {Rahm},\ and\ \citenamefont {Holze}}]{Wu2003}%
  \BibitemOpen
  \bibfield  {author} {\bibinfo {author} {\bibfnamefont {Y.}~\bibnamefont
  {Wu}}, \bibinfo {author} {\bibfnamefont {E.}~\bibnamefont {Rahm}}, \ and\
  \bibinfo {author} {\bibfnamefont {R.}~\bibnamefont {Holze}},\ }\href
  {\doibase 10.1016/S0378-7753(02)00596-7} {\bibfield  {journal} {\bibinfo
  {journal} {J. Power Sources}\ }\textbf {\bibinfo {volume} {114}},\ \bibinfo
  {pages} {228} (\bibinfo {year} {2003})}\BibitemShut {NoStop}%
\bibitem [{\citenamefont {Kaskhedikar}\ and\ \citenamefont
  {Maier}(2009)}]{Kaskhedikar2009}%
  \BibitemOpen
  \bibfield  {author} {\bibinfo {author} {\bibfnamefont {N.~A.}\ \bibnamefont
  {Kaskhedikar}}\ and\ \bibinfo {author} {\bibfnamefont {J.}~\bibnamefont
  {Maier}},\ }\href {\doibase 10.1002/adma.200901079} {\bibfield  {journal}
  {\bibinfo  {journal} {Adv. Mater.}\ }\textbf {\bibinfo {volume} {21}},\
  \bibinfo {pages} {2664} (\bibinfo {year} {2009})}\BibitemShut {NoStop}%
\bibitem [{\citenamefont {Etacheri}\ \emph {et~al.}(2011)\citenamefont
  {Etacheri}, \citenamefont {Marom}, \citenamefont {Elazari}, \citenamefont
  {Salitra},\ and\ \citenamefont {Aurbach}}]{Etacheri2011}%
  \BibitemOpen
  \bibfield  {author} {\bibinfo {author} {\bibfnamefont {V.}~\bibnamefont
  {Etacheri}}, \bibinfo {author} {\bibfnamefont {R.}~\bibnamefont {Marom}},
  \bibinfo {author} {\bibfnamefont {R.}~\bibnamefont {Elazari}}, \bibinfo
  {author} {\bibfnamefont {G.}~\bibnamefont {Salitra}}, \ and\ \bibinfo
  {author} {\bibfnamefont {D.}~\bibnamefont {Aurbach}},\ }\href {\doibase
  10.1039/C1EE01598B} {\bibfield  {journal} {\bibinfo  {journal} {Energy
  Environ. Sci.}\ }\textbf {\bibinfo {volume} {4}},\ \bibinfo {pages} {3243}
  (\bibinfo {year} {2011})}\BibitemShut {NoStop}%
\bibitem [{\citenamefont {Nishimura}\ \emph {et~al.}(2008)\citenamefont
  {Nishimura}, \citenamefont {Kobayashi}, \citenamefont {Ohoyama},
  \citenamefont {Kanno}, \citenamefont {Yashima},\ and\ \citenamefont
  {Yamada}}]{Nishimura2008}%
  \BibitemOpen
  \bibfield  {author} {\bibinfo {author} {\bibfnamefont {S.-i.}\ \bibnamefont
  {Nishimura}}, \bibinfo {author} {\bibfnamefont {G.}~\bibnamefont
  {Kobayashi}}, \bibinfo {author} {\bibfnamefont {K.}~\bibnamefont {Ohoyama}},
  \bibinfo {author} {\bibfnamefont {R.}~\bibnamefont {Kanno}}, \bibinfo
  {author} {\bibfnamefont {M.}~\bibnamefont {Yashima}}, \ and\ \bibinfo
  {author} {\bibfnamefont {A.}~\bibnamefont {Yamada}},\ }\href
  {http://dx.doi.org/10.1038/nmat2251
  http://www.nature.com/nmat/journal/v7/n9/suppinfo/nmat2251{\_}S1.html}
  {\bibfield  {journal} {\bibinfo  {journal} {Nat. Mater.}\ }\textbf {\bibinfo
  {volume} {7}},\ \bibinfo {pages} {707} (\bibinfo {year} {2008})}\BibitemShut
  {NoStop}%
\bibitem [{\citenamefont {Balke}\ \emph {et~al.}(2010)\citenamefont {Balke},
  \citenamefont {Jesse}, \citenamefont {Morozovska}, \citenamefont {Eliseev},
  \citenamefont {Chung}, \citenamefont {Kim}, \citenamefont {Adamczyk},
  \citenamefont {Garcia}, \citenamefont {Dudney},\ and\ \citenamefont
  {Kalinin}}]{Balke2010}%
  \BibitemOpen
  \bibfield  {author} {\bibinfo {author} {\bibfnamefont {N.}~\bibnamefont
  {Balke}}, \bibinfo {author} {\bibfnamefont {S.}~\bibnamefont {Jesse}},
  \bibinfo {author} {\bibfnamefont {A.~N.}\ \bibnamefont {Morozovska}},
  \bibinfo {author} {\bibfnamefont {E.}~\bibnamefont {Eliseev}}, \bibinfo
  {author} {\bibfnamefont {D.~W.}\ \bibnamefont {Chung}}, \bibinfo {author}
  {\bibfnamefont {Y.}~\bibnamefont {Kim}}, \bibinfo {author} {\bibfnamefont
  {L.}~\bibnamefont {Adamczyk}}, \bibinfo {author} {\bibfnamefont {R.~E.}\
  \bibnamefont {Garcia}}, \bibinfo {author} {\bibfnamefont {N.}~\bibnamefont
  {Dudney}}, \ and\ \bibinfo {author} {\bibfnamefont {S.~V.}\ \bibnamefont
  {Kalinin}},\ }\href {http://dx.doi.org/10.1038/nnano.2010.174
  http://www.nature.com/nnano/journal/v5/n10/abs/nnano.2010.174.html{\#}supplementary-information}
  {\bibfield  {journal} {\bibinfo  {journal} {Nat. Nanotechnol.}\ }\textbf
  {\bibinfo {volume} {5}},\ \bibinfo {pages} {749} (\bibinfo {year}
  {2010})}\BibitemShut {NoStop}%
\bibitem [{\citenamefont {Pecher}\ \emph {et~al.}(2017)\citenamefont {Pecher},
  \citenamefont {Carretero-Gonz{\'{a}}lez}, \citenamefont {Griffith},\ and\
  \citenamefont {Grey}}]{Pecher2016}%
  \BibitemOpen
  \bibfield  {author} {\bibinfo {author} {\bibfnamefont {O.}~\bibnamefont
  {Pecher}}, \bibinfo {author} {\bibfnamefont {J.}~\bibnamefont
  {Carretero-Gonz{\'{a}}lez}}, \bibinfo {author} {\bibfnamefont {K.~J.}\
  \bibnamefont {Griffith}}, \ and\ \bibinfo {author} {\bibfnamefont {C.~P.}\
  \bibnamefont {Grey}},\ }\href {\doibase 10.1021/acs.chemmater.6b03183}
  {\bibfield  {journal} {\bibinfo  {journal} {Chem. Mater.}\ }\textbf {\bibinfo
  {volume} {29}},\ \bibinfo {pages} {213} (\bibinfo {year} {2017})}\BibitemShut
  {NoStop}%
\bibitem [{\citenamefont {Ceder}\ \emph {et~al.}(1998)\citenamefont {Ceder},
  \citenamefont {Chiang}, \citenamefont {Sadoway}, \citenamefont {Aydinol},
  \citenamefont {Jang},\ and\ \citenamefont {Huang}}]{Ceder1998}%
  \BibitemOpen
  \bibfield  {author} {\bibinfo {author} {\bibfnamefont {G.}~\bibnamefont
  {Ceder}}, \bibinfo {author} {\bibfnamefont {Y.-M.}\ \bibnamefont {Chiang}},
  \bibinfo {author} {\bibfnamefont {D.~R.}\ \bibnamefont {Sadoway}}, \bibinfo
  {author} {\bibfnamefont {M.~K.}\ \bibnamefont {Aydinol}}, \bibinfo {author}
  {\bibfnamefont {Y.-I.}\ \bibnamefont {Jang}}, \ and\ \bibinfo {author}
  {\bibfnamefont {B.}~\bibnamefont {Huang}},\ }\href
  {http://dx.doi.org/10.1038/33647} {\bibfield  {journal} {\bibinfo  {journal}
  {Nature}\ }\textbf {\bibinfo {volume} {392}},\ \bibinfo {pages} {694}
  (\bibinfo {year} {1998})}\BibitemShut {NoStop}%
\bibitem [{\citenamefont {Meng}\ and\ \citenamefont {{Arroyo-de
  Dompablo}}(2009)}]{Meng2009}%
  \BibitemOpen
  \bibfield  {author} {\bibinfo {author} {\bibfnamefont {Y.~S.}\ \bibnamefont
  {Meng}}\ and\ \bibinfo {author} {\bibfnamefont {M.~E.}\ \bibnamefont
  {{Arroyo-de Dompablo}}},\ }\href {\doibase 10.1039/B901825E} {\bibfield
  {journal} {\bibinfo  {journal} {Energy Environ. Sci.}\ }\textbf {\bibinfo
  {volume} {2}},\ \bibinfo {pages} {589} (\bibinfo {year} {2009})}\BibitemShut
  {NoStop}%
\bibitem [{\citenamefont {Islam}\ and\ \citenamefont
  {Fisher}(2014)}]{Islam2014}%
  \BibitemOpen
  \bibfield  {author} {\bibinfo {author} {\bibfnamefont {M.~S.}\ \bibnamefont
  {Islam}}\ and\ \bibinfo {author} {\bibfnamefont {C.~A.~J.}\ \bibnamefont
  {Fisher}},\ }\href {\doibase 10.1039/C3CS60199D} {\bibfield  {journal}
  {\bibinfo  {journal} {Chem. Soc. Rev.}\ }\textbf {\bibinfo {volume} {43}},\
  \bibinfo {pages} {185} (\bibinfo {year} {2014})}\BibitemShut {NoStop}%
\bibitem [{\citenamefont {Stratford}\ \emph {et~al.}(2017)\citenamefont
  {Stratford}, \citenamefont {Mayo}, \citenamefont {Allan}, \citenamefont
  {Pecher}, \citenamefont {Borkiewicz}, \citenamefont {Wiaderek}, \citenamefont
  {Chapman}, \citenamefont {Pickard}, \citenamefont {Morris},\ and\
  \citenamefont {Grey}}]{Stratford2017}%
  \BibitemOpen
  \bibfield  {author} {\bibinfo {author} {\bibfnamefont {J.~M.}\ \bibnamefont
  {Stratford}}, \bibinfo {author} {\bibfnamefont {M.}~\bibnamefont {Mayo}},
  \bibinfo {author} {\bibfnamefont {P.~K.}\ \bibnamefont {Allan}}, \bibinfo
  {author} {\bibfnamefont {O.}~\bibnamefont {Pecher}}, \bibinfo {author}
  {\bibfnamefont {O.~J.}\ \bibnamefont {Borkiewicz}}, \bibinfo {author}
  {\bibfnamefont {K.~M.}\ \bibnamefont {Wiaderek}}, \bibinfo {author}
  {\bibfnamefont {K.~W.}\ \bibnamefont {Chapman}}, \bibinfo {author}
  {\bibfnamefont {C.~J.}\ \bibnamefont {Pickard}}, \bibinfo {author}
  {\bibfnamefont {A.~J.}\ \bibnamefont {Morris}}, \ and\ \bibinfo {author}
  {\bibfnamefont {C.~P.}\ \bibnamefont {Grey}},\ }\href {\doibase
  10.1021/jacs.7b01398} {\bibfield  {journal} {\bibinfo  {journal} {J. Am.
  Chem. Soc.}\ }\textbf {\bibinfo {volume} {139}},\ \bibinfo {pages} {7273} (\bibinfo {year}
  {2017})}\BibitemShut {NoStop}%
\bibitem [{\citenamefont {Kganyago}\ and\ \citenamefont
  {Ngoepe}(2003)}]{Kganyago2003}%
  \BibitemOpen
  \bibfield  {author} {\bibinfo {author} {\bibfnamefont {K.~R.}\ \bibnamefont
  {Kganyago}}\ and\ \bibinfo {author} {\bibfnamefont {P.~E.}\ \bibnamefont
  {Ngoepe}},\ }\href {\doibase 10.1103/PhysRevB.68.205111} {\bibfield
  {journal} {\bibinfo  {journal} {Phys. Rev. B}\ }\textbf {\bibinfo {volume}
  {68}},\ \bibinfo {pages} {205111} (\bibinfo {year} {2003})}\BibitemShut
  {NoStop}%
\bibitem [{\citenamefont {Toyoura}\ \emph {et~al.}(2008)\citenamefont
  {Toyoura}, \citenamefont {Koyama}, \citenamefont {Kuwabara}, \citenamefont
  {Oba},\ and\ \citenamefont {Tanaka}}]{Toyoura2008}%
  \BibitemOpen
  \bibfield  {author} {\bibinfo {author} {\bibfnamefont {K.}~\bibnamefont
  {Toyoura}}, \bibinfo {author} {\bibfnamefont {Y.}~\bibnamefont {Koyama}},
  \bibinfo {author} {\bibfnamefont {A.}~\bibnamefont {Kuwabara}}, \bibinfo
  {author} {\bibfnamefont {F.}~\bibnamefont {Oba}}, \ and\ \bibinfo {author}
  {\bibfnamefont {I.}~\bibnamefont {Tanaka}},\ }\href
  {https://link.aps.org/doi/10.1103/PhysRevB.78.214303} {\bibfield  {journal}
  {\bibinfo  {journal} {Phys. Rev. B}\ }\textbf {\bibinfo {volume} {78}},\
  \bibinfo {pages} {214303} (\bibinfo {year} {2008})}\BibitemShut {NoStop}%
\bibitem [{\citenamefont {Khantha}\ \emph {et~al.}(2004)\citenamefont
  {Khantha}, \citenamefont {Cordero}, \citenamefont {Molina}, \citenamefont
  {Alonso},\ and\ \citenamefont {Girifalco}}]{Khantha2004}%
  \BibitemOpen
  \bibfield  {author} {\bibinfo {author} {\bibfnamefont {M.}~\bibnamefont
  {Khantha}}, \bibinfo {author} {\bibfnamefont {N.~A.}\ \bibnamefont
  {Cordero}}, \bibinfo {author} {\bibfnamefont {L.~M.}\ \bibnamefont {Molina}},
  \bibinfo {author} {\bibfnamefont {J.~A.}\ \bibnamefont {Alonso}}, \ and\
  \bibinfo {author} {\bibfnamefont {L.~A.}\ \bibnamefont {Girifalco}},\ }\href
  {\doibase 10.1103/PhysRevB.70.125422} {\bibfield  {journal} {\bibinfo
  {journal} {Phys. Rev. B}\ }\textbf {\bibinfo {volume} {70}},\ \bibinfo
  {pages} {125422} (\bibinfo {year} {2004})}\BibitemShut {NoStop}%
\bibitem [{\citenamefont {Rytk{\"{o}}nen}\ \emph {et~al.}(2007)\citenamefont
  {Rytk{\"{o}}nen}, \citenamefont {Akola},\ and\ \citenamefont
  {Manninen}}]{Rytkonen2007}%
  \BibitemOpen
  \bibfield  {author} {\bibinfo {author} {\bibfnamefont {K.}~\bibnamefont
  {Rytk{\"{o}}nen}}, \bibinfo {author} {\bibfnamefont {J.}~\bibnamefont
  {Akola}}, \ and\ \bibinfo {author} {\bibfnamefont {M.}~\bibnamefont
  {Manninen}},\ }\href {\doibase 10.1103/PhysRevB.75.075401} {\bibfield
  {journal} {\bibinfo  {journal} {Phys. Rev. B}\ }\textbf {\bibinfo {volume}
  {75}},\ \bibinfo {pages} {075401} (\bibinfo {year} {2007})}\BibitemShut
  {NoStop}%
\bibitem [{\citenamefont {Persson}\ \emph {et~al.}(2010)\citenamefont
  {Persson}, \citenamefont {Sethuraman}, \citenamefont {Hardwick},
  \citenamefont {Hinuma}, \citenamefont {Meng}, \citenamefont {van~der Ven},
  \citenamefont {Srinivasan}, \citenamefont {Kostecki},\ and\ \citenamefont
  {Ceder}}]{Persson2010}%
  \BibitemOpen
  \bibfield  {author} {\bibinfo {author} {\bibfnamefont {K.}~\bibnamefont
  {Persson}}, \bibinfo {author} {\bibfnamefont {V.~A.}\ \bibnamefont
  {Sethuraman}}, \bibinfo {author} {\bibfnamefont {L.~J.}\ \bibnamefont
  {Hardwick}}, \bibinfo {author} {\bibfnamefont {Y.}~\bibnamefont {Hinuma}},
  \bibinfo {author} {\bibfnamefont {Y.~S.}\ \bibnamefont {Meng}}, \bibinfo
  {author} {\bibfnamefont {A.}~\bibnamefont {van~der Ven}}, \bibinfo {author}
  {\bibfnamefont {V.}~\bibnamefont {Srinivasan}}, \bibinfo {author}
  {\bibfnamefont {R.}~\bibnamefont {Kostecki}}, \ and\ \bibinfo {author}
  {\bibfnamefont {G.}~\bibnamefont {Ceder}},\ }\href {\doibase
  10.1021/jz100188d} {\bibfield  {journal} {\bibinfo  {journal} {J. Phys. Chem.
  Lett.}\ }\textbf {\bibinfo {volume} {1}},\ \bibinfo {pages} {1176} (\bibinfo
  {year} {2010})}\BibitemShut {NoStop}%
\bibitem [{\citenamefont {Fan}\ \emph {et~al.}(2012)\citenamefont {Fan},
  \citenamefont {Zheng},\ and\ \citenamefont {Kuo}}]{Fan2012}%
  \BibitemOpen
  \bibfield  {author} {\bibinfo {author} {\bibfnamefont {X.}~\bibnamefont
  {Fan}}, \bibinfo {author} {\bibfnamefont {W.~T.}\ \bibnamefont {Zheng}}, \
  and\ \bibinfo {author} {\bibfnamefont {J.-L.}\ \bibnamefont {Kuo}},\
  }\href@noop {} {\bibfield  {journal} {\bibinfo  {journal} {ACS Appl. Mater.
  Interfaces}\ }\textbf {\bibinfo {volume} {4}},\ \bibinfo {pages} {2432}
  (\bibinfo {year} {2012})}\BibitemShut {NoStop}%
\bibitem [{\citenamefont {Lee}\ and\ \citenamefont {Persson}(2012)}]{Lee2012a}%
  \BibitemOpen
  \bibfield  {author} {\bibinfo {author} {\bibfnamefont {E.}~\bibnamefont
  {Lee}}\ and\ \bibinfo {author} {\bibfnamefont {K.~A.}\ \bibnamefont
  {Persson}},\ }\href {\doibase 10.1021/nl3019164} {\bibfield  {journal}
  {\bibinfo  {journal} {Nano Lett.}\ }\textbf {\bibinfo {volume} {12}},\
  \bibinfo {pages} {4624} (\bibinfo {year} {2012})}\BibitemShut {NoStop}%
\bibitem [{\citenamefont {Zhou}\ \emph {et~al.}(2012)\citenamefont {Zhou},
  \citenamefont {Hou},\ and\ \citenamefont {Wu}}]{Zhou2012}%
  \BibitemOpen
  \bibfield  {author} {\bibinfo {author} {\bibfnamefont {L.-J.}\ \bibnamefont
  {Zhou}}, \bibinfo {author} {\bibfnamefont {Z.~F.}\ \bibnamefont {Hou}}, \
  and\ \bibinfo {author} {\bibfnamefont {L.-M.}\ \bibnamefont {Wu}},\ }\href
  {\doibase 10.1021/jp304861d} {\bibfield  {journal} {\bibinfo  {journal} {J.
  Phys. Chem. C}\ }\textbf {\bibinfo {volume} {116}},\ \bibinfo {pages} {21780}
  (\bibinfo {year} {2012})}\BibitemShut {NoStop}%
\bibitem [{\citenamefont {Liu}\ \emph {et~al.}(2013)\citenamefont {Liu},
  \citenamefont {Artyukhov}, \citenamefont {Liu}, \citenamefont {Harutyunyan},\
  and\ \citenamefont {Yakobson}}]{Liu2013a}%
  \BibitemOpen
  \bibfield  {author} {\bibinfo {author} {\bibfnamefont {Y.}~\bibnamefont
  {Liu}}, \bibinfo {author} {\bibfnamefont {V.~I.}\ \bibnamefont {Artyukhov}},
  \bibinfo {author} {\bibfnamefont {M.}~\bibnamefont {Liu}}, \bibinfo {author}
  {\bibfnamefont {A.~R.}\ \bibnamefont {Harutyunyan}}, \ and\ \bibinfo {author}
  {\bibfnamefont {B.~I.}\ \bibnamefont {Yakobson}},\ }\href {\doibase
  10.1021/jz400491b} {\bibfield  {journal} {\bibinfo  {journal} {J. Phys. Chem.
  Lett.}\ }\textbf {\bibinfo {volume} {4}},\ \bibinfo {pages} {1737} (\bibinfo
  {year} {2013})}\BibitemShut {NoStop}%
\bibitem [{\citenamefont {Liu}\ \emph {et~al.}(2014)\citenamefont {Liu},
  \citenamefont {Kutana}, \citenamefont {Liu},\ and\ \citenamefont
  {Yakobson}}]{Liu2014b}%
  \BibitemOpen
  \bibfield  {author} {\bibinfo {author} {\bibfnamefont {M.}~\bibnamefont
  {Liu}}, \bibinfo {author} {\bibfnamefont {A.}~\bibnamefont {Kutana}},
  \bibinfo {author} {\bibfnamefont {Y.}~\bibnamefont {Liu}}, \ and\ \bibinfo
  {author} {\bibfnamefont {B.~I.}\ \bibnamefont {Yakobson}},\ }\href {\doibase
  10.1021/jz500199d} {\bibfield  {journal} {\bibinfo  {journal} {J. Phys. Chem.
  Lett.}\ }\textbf {\bibinfo {volume} {5}},\ \bibinfo {pages} {1225} (\bibinfo
  {year} {2014})}\BibitemShut {NoStop}%
\bibitem [{\citenamefont {Li}\ and\ \citenamefont {Merz}(2017)}]{Li2017}%
  \BibitemOpen
  \bibfield  {author} {\bibinfo {author} {\bibfnamefont {P.}~\bibnamefont
  {Li}}\ and\ \bibinfo {author} {\bibfnamefont {K.~M.}\ \bibnamefont {Merz}},\
  }\href {\doibase 10.1021/acs.chemrev.6b00440} {\bibfield  {journal} {\bibinfo
   {journal} {Chem. Rev.}\ }\textbf {\bibinfo {volume} {117}},\ \bibinfo
  {pages} {1564} (\bibinfo {year} {2017})}\BibitemShut {NoStop}%
\bibitem [{\citenamefont {Han}\ \emph {et~al.}(2005)\citenamefont {Han},
  \citenamefont {van Duin}, \citenamefont {Goddard},\ and\ \citenamefont
  {Lee}}]{Han2005}%
  \BibitemOpen
  \bibfield  {author} {\bibinfo {author} {\bibfnamefont {S.~S.}\ \bibnamefont
  {Han}}, \bibinfo {author} {\bibfnamefont {A.~C.~T.}\ \bibnamefont {van
  Duin}}, \bibinfo {author} {\bibfnamefont {W.~A.}\ \bibnamefont {Goddard}}, \
  and\ \bibinfo {author} {\bibfnamefont {H.~M.}\ \bibnamefont {Lee}},\ }\href
  {\doibase 10.1021/jp051450m} {\bibfield  {journal} {\bibinfo  {journal} {J.
  Phys. Chem. A}\ }\textbf {\bibinfo {volume} {109}},\ \bibinfo {pages} {4575}
  (\bibinfo {year} {2005})}\BibitemShut {NoStop}%
\bibitem [{\citenamefont {Yang}\ \emph {et~al.}(2013)\citenamefont {Yang},
  \citenamefont {Huang}, \citenamefont {Liang}, \citenamefont {van Duin},
  \citenamefont {Raju},\ and\ \citenamefont {Zhang}}]{Yang2013}%
  \BibitemOpen
  \bibfield  {author} {\bibinfo {author} {\bibfnamefont {H.}~\bibnamefont
  {Yang}}, \bibinfo {author} {\bibfnamefont {X.}~\bibnamefont {Huang}},
  \bibinfo {author} {\bibfnamefont {W.}~\bibnamefont {Liang}}, \bibinfo
  {author} {\bibfnamefont {A.~C.~T.}\ \bibnamefont {van Duin}}, \bibinfo
  {author} {\bibfnamefont {M.}~\bibnamefont {Raju}}, \ and\ \bibinfo {author}
  {\bibfnamefont {S.}~\bibnamefont {Zhang}},\ }\href {\doibase
  https://doi.org/10.1016/j.cplett.2013.01.048} {\bibfield  {journal} {\bibinfo
   {journal} {Chem. Phys. Lett.}\ }\textbf {\bibinfo {volume} {563}},\ \bibinfo
  {pages} {58} (\bibinfo {year} {2013})}\BibitemShut {NoStop}%
\bibitem [{\citenamefont {Huang}\ \emph {et~al.}(2013)\citenamefont {Huang},
  \citenamefont {Yang}, \citenamefont {Liang}, \citenamefont {Raju},
  \citenamefont {Terrones}, \citenamefont {Crespi}, \citenamefont {{Van
  Duin}},\ and\ \citenamefont {Zhang}}]{Huang2013}%
  \BibitemOpen
  \bibfield  {author} {\bibinfo {author} {\bibfnamefont {X.}~\bibnamefont
  {Huang}}, \bibinfo {author} {\bibfnamefont {H.}~\bibnamefont {Yang}},
  \bibinfo {author} {\bibfnamefont {W.}~\bibnamefont {Liang}}, \bibinfo
  {author} {\bibfnamefont {M.}~\bibnamefont {Raju}}, \bibinfo {author}
  {\bibfnamefont {M.}~\bibnamefont {Terrones}}, \bibinfo {author}
  {\bibfnamefont {V.~H.}\ \bibnamefont {Crespi}}, \bibinfo {author}
  {\bibfnamefont {A.~C.~T.}\ \bibnamefont {{Van Duin}}}, \ and\ \bibinfo
  {author} {\bibfnamefont {S.}~\bibnamefont {Zhang}},\ }\href {\doibase
  10.1063/1.4824418} {\bibfield  {journal} {\bibinfo  {journal} {Appl. Phys.
  Lett.}\ }\textbf {\bibinfo {volume} {103}},\ \bibinfo {pages} {153901}
  (\bibinfo {year} {2013})}\BibitemShut {NoStop}%
\bibitem [{\citenamefont {Raju}\ \emph {et~al.}(2015)\citenamefont {Raju},
  \citenamefont {Ganesh}, \citenamefont {Kent},\ and\ \citenamefont {van
  Duin}}]{Raju2015}%
  \BibitemOpen
  \bibfield  {author} {\bibinfo {author} {\bibfnamefont {M.}~\bibnamefont
  {Raju}}, \bibinfo {author} {\bibfnamefont {P.}~\bibnamefont {Ganesh}},
  \bibinfo {author} {\bibfnamefont {P.~R.~C.}\ \bibnamefont {Kent}}, \ and\
  \bibinfo {author} {\bibfnamefont {A.~C.~T.}\ \bibnamefont {van Duin}},\
  }\href {\doibase 10.1021/ct501027v} {\bibfield  {journal} {\bibinfo
  {journal} {J. Chem. Theory Comput.}\ }\textbf {\bibinfo {volume} {11}},\
  \bibinfo {pages} {2156} (\bibinfo {year} {2015})}\BibitemShut {NoStop}%
\bibitem [{\citenamefont {McNutt}\ \emph {et~al.}(2017)\citenamefont {McNutt},
  \citenamefont {McDonnell}, \citenamefont {Rios},\ and\ \citenamefont
  {Keffer}}]{McNutt2017}%
  \BibitemOpen
  \bibfield  {author} {\bibinfo {author} {\bibfnamefont {N.~W.}\ \bibnamefont
  {McNutt}}, \bibinfo {author} {\bibfnamefont {M.~T.}\ \bibnamefont
  {McDonnell}}, \bibinfo {author} {\bibfnamefont {O.}~\bibnamefont {Rios}}, \
  and\ \bibinfo {author} {\bibfnamefont {D.~J.}\ \bibnamefont {Keffer}},\
  }\href {\doibase 10.1021/acsami.6b13748} {\bibfield  {journal} {\bibinfo
  {journal} {ACS Appl. Mater. Interfaces}\ }\textbf {\bibinfo {volume} {9}},\
  \bibinfo {pages} {6988} (\bibinfo {year} {2017})}\BibitemShut {NoStop}%
\bibitem [{\citenamefont {Pastewka}\ \emph {et~al.}(2008)\citenamefont
  {Pastewka}, \citenamefont {Pou}, \citenamefont {P{\'{e}}rez}, \citenamefont
  {Gumbsch},\ and\ \citenamefont {Moseler}}]{Pastewka2008}%
  \BibitemOpen
  \bibfield  {author} {\bibinfo {author} {\bibfnamefont {L.}~\bibnamefont
  {Pastewka}}, \bibinfo {author} {\bibfnamefont {P.}~\bibnamefont {Pou}},
  \bibinfo {author} {\bibfnamefont {R.}~\bibnamefont {P{\'{e}}rez}}, \bibinfo
  {author} {\bibfnamefont {P.}~\bibnamefont {Gumbsch}}, \ and\ \bibinfo
  {author} {\bibfnamefont {M.}~\bibnamefont {Moseler}},\ }\href {\doibase
  10.1103/PhysRevB.78.161402} {\bibfield  {journal} {\bibinfo  {journal}
  {Physical Review B}\ }\textbf {\bibinfo {volume} {78}},\ \bibinfo {pages}
  {161402} (\bibinfo {year} {2008})}\BibitemShut {NoStop}%
\bibitem [{\citenamefont {Pastewka}\ \emph {et~al.}(2012)\citenamefont
  {Pastewka}, \citenamefont {Mrovec}, \citenamefont {Moseler},\ and\
  \citenamefont {Gumbsch}}]{Pastewka2012}%
  \BibitemOpen
  \bibfield  {author} {\bibinfo {author} {\bibfnamefont {L.}~\bibnamefont
  {Pastewka}}, \bibinfo {author} {\bibfnamefont {M.}~\bibnamefont {Mrovec}},
  \bibinfo {author} {\bibfnamefont {M.}~\bibnamefont {Moseler}}, \ and\
  \bibinfo {author} {\bibfnamefont {P.}~\bibnamefont {Gumbsch}},\ }\href
  {\doibase DOI: 10.1557/mrs.2012.94} {\bibfield  {journal} {\bibinfo
  {journal} {MRS Bull.}\ }\textbf {\bibinfo {volume} {37}},\ \bibinfo {pages}
  {493} (\bibinfo {year} {2012})}\BibitemShut {NoStop}%
\bibitem [{\citenamefont {de~Tomas}\ \emph {et~al.}(2016)\citenamefont
  {de~Tomas}, \citenamefont {Suarez-Martinez},\ and\ \citenamefont
  {Marks}}]{deTomas2016}%
  \BibitemOpen
  \bibfield  {author} {\bibinfo {author} {\bibfnamefont {C.}~\bibnamefont
  {de~Tomas}}, \bibinfo {author} {\bibfnamefont {I.}~\bibnamefont
  {Suarez-Martinez}}, \ and\ \bibinfo {author} {\bibfnamefont {N.~A.}\
  \bibnamefont {Marks}},\ }\href {\doibase 10.1016/j.carbon.2016.08.024}
  {\bibfield  {journal} {\bibinfo  {journal} {Carbon}\ }\textbf {\bibinfo
  {volume} {109}},\ \bibinfo {pages} {681} (\bibinfo {year}
  {2016})}\BibitemShut {NoStop}%
\bibitem [{\citenamefont {Behler}\ and\ \citenamefont
  {Parrinello}(2007)}]{Behler2007}%
  \BibitemOpen
  \bibfield  {author} {\bibinfo {author} {\bibfnamefont {J.}~\bibnamefont
  {Behler}}\ and\ \bibinfo {author} {\bibfnamefont {M.}~\bibnamefont
  {Parrinello}},\ }\href {\doibase 10.1103/PhysRevLett.98.146401} {\bibfield
  {journal} {\bibinfo  {journal} {Phys. Rev. Lett.}\ }\textbf {\bibinfo
  {volume} {98}},\ \bibinfo {pages} {146401} (\bibinfo {year}
  {2007})}\BibitemShut {NoStop}%
\bibitem [{\citenamefont {Artrith}\ \emph {et~al.}(2011)\citenamefont
  {Artrith}, \citenamefont {Morawietz},\ and\ \citenamefont
  {Behler}}]{Artrith2011}%
  \BibitemOpen
  \bibfield  {author} {\bibinfo {author} {\bibfnamefont {N.}~\bibnamefont
  {Artrith}}, \bibinfo {author} {\bibfnamefont {T.}~\bibnamefont {Morawietz}},
  \ and\ \bibinfo {author} {\bibfnamefont {J.}~\bibnamefont {Behler}},\ }\href
  {\doibase 10.1103/PhysRevB.83.153101} {\bibfield  {journal} {\bibinfo
  {journal} {Phys. Rev. B}\ }\textbf {\bibinfo {volume} {83}},\ \bibinfo
  {pages} {153101} (\bibinfo {year} {2011})}\BibitemShut {NoStop}%
\bibitem [{\citenamefont {Sosso}\ \emph {et~al.}(2012)\citenamefont {Sosso},
  \citenamefont {Miceli}, \citenamefont {Caravati}, \citenamefont {Behler},\
  and\ \citenamefont {Bernasconi}}]{Sosso2012}%
  \BibitemOpen
  \bibfield  {author} {\bibinfo {author} {\bibfnamefont {G.~C.}\ \bibnamefont
  {Sosso}}, \bibinfo {author} {\bibfnamefont {G.}~\bibnamefont {Miceli}},
  \bibinfo {author} {\bibfnamefont {S.}~\bibnamefont {Caravati}}, \bibinfo
  {author} {\bibfnamefont {J.}~\bibnamefont {Behler}}, \ and\ \bibinfo {author}
  {\bibfnamefont {M.}~\bibnamefont {Bernasconi}},\ }\href {\doibase
  10.1103/PhysRevB.85.174103} {\bibfield  {journal} {\bibinfo  {journal} {Phys.
  Rev. B}\ }\textbf {\bibinfo {volume} {85}},\ \bibinfo {pages} {174103}
  (\bibinfo {year} {2012})}\BibitemShut {NoStop}%
\bibitem [{\citenamefont {Artrith}\ and\ \citenamefont
  {Urban}(2016)}]{Artrith2016}%
  \BibitemOpen
  \bibfield  {author} {\bibinfo {author} {\bibfnamefont {N.}~\bibnamefont
  {Artrith}}\ and\ \bibinfo {author} {\bibfnamefont {A.}~\bibnamefont
  {Urban}},\ }\href {\doibase 10.1016/j.commatsci.2015.11.047} {\bibfield
  {journal} {\bibinfo  {journal} {Comput. Mater. Sci.}\ }\textbf {\bibinfo
  {volume} {114}},\ \bibinfo {pages} {135} (\bibinfo {year}
  {2016})}\BibitemShut {NoStop}%
\bibitem{Smith2017}
   J.~S. Smith, O.~Isayev, and A.~E. Roitberg, 
   Chem. Sci. {\bf 8}, 3192 (2017).
\bibitem [{\citenamefont {Hajinazar}\ \emph {et~al.}(2017)\citenamefont
  {Hajinazar}, \citenamefont {Shao},\ and\ \citenamefont
  {Kolmogorov}}]{Hajinazar2017}%
  \BibitemOpen
  \bibfield  {author} {\bibinfo {author} {\bibfnamefont {S.}~\bibnamefont
  {Hajinazar}}, \bibinfo {author} {\bibfnamefont {J.}~\bibnamefont {Shao}}, \
  and\ \bibinfo {author} {\bibfnamefont {A.~N.}\ \bibnamefont {Kolmogorov}},\
  }\href {\doibase 10.1103/PhysRevB.95.014114} {\bibfield  {journal} {\bibinfo
  {journal} {Phys. Rev. B}\ }\textbf {\bibinfo {volume} {95}},\ \bibinfo
  {pages} {14114} (\bibinfo {year} {2017})}\BibitemShut {NoStop}%
\bibitem [{\citenamefont {Faraji}\ \emph {et~al.}(2017)\citenamefont {Faraji},
  \citenamefont {Ghasemi}, \citenamefont {Rostami}, \citenamefont
  {Rasoulkhani}, \citenamefont {Schaefer}, \citenamefont {Goedecker},\ and\
  \citenamefont {Amsler}}]{Faraji2017}%
  \BibitemOpen
  \bibfield  {author} {\bibinfo {author} {\bibfnamefont {S.}~\bibnamefont
  {Faraji}}, \bibinfo {author} {\bibfnamefont {S.~A.}\ \bibnamefont {Ghasemi}},
  \bibinfo {author} {\bibfnamefont {S.}~\bibnamefont {Rostami}}, \bibinfo
  {author} {\bibfnamefont {R.}~\bibnamefont {Rasoulkhani}}, \bibinfo {author}
  {\bibfnamefont {B.}~\bibnamefont {Schaefer}}, \bibinfo {author}
  {\bibfnamefont {S.}~\bibnamefont {Goedecker}}, \ and\ \bibinfo {author}
  {\bibfnamefont {M.}~\bibnamefont {Amsler}},\ }\href {\doibase
  10.1103/PhysRevB.95.104105} {\bibfield  {journal} {\bibinfo  {journal} {Phys.
  Rev. B}\ }\textbf {\bibinfo {volume} {95}},\ \bibinfo {pages} {104105}
  (\bibinfo {year} {2017})}\BibitemShut {NoStop}%
\bibitem{Kobayashi2017} 
  R.~Kobayashi, D.~Giofr\'e{}, T.~Junge, M.~Ceriotti, and W.~A. Curtin,
  Phys. Rev. Mater. {\bf 1}, 053604 (2017).
\bibitem [{\citenamefont {Bart{\'{o}}k}\ \emph {et~al.}(2010)\citenamefont
  {Bart{\'{o}}k}, \citenamefont {Payne}, \citenamefont {Kondor},\ and\
  \citenamefont {Cs{\'{a}}nyi}}]{Bartok2010}%
  \BibitemOpen
  \bibfield  {author} {\bibinfo {author} {\bibfnamefont {A.~P.}\ \bibnamefont
  {Bart{\'{o}}k}}, \bibinfo {author} {\bibfnamefont {M.~C.}\ \bibnamefont
  {Payne}}, \bibinfo {author} {\bibfnamefont {R.}~\bibnamefont {Kondor}}, \
  and\ \bibinfo {author} {\bibfnamefont {G.}~\bibnamefont {Cs{\'{a}}nyi}},\
  }\href@noop {} {\bibfield  {journal} {\bibinfo  {journal} {Phys. Rev. Lett.}\
  }\textbf {\bibinfo {volume} {104}},\ \bibinfo {pages} {136403} (\bibinfo
  {year} {2010})}\BibitemShut {NoStop}%
\bibitem [{\citenamefont {Szlachta}\ \emph {et~al.}(2014)\citenamefont
  {Szlachta}, \citenamefont {Bart{\'{o}}k},\ and\ \citenamefont
  {Cs{\'{a}}nyi}}]{Szlachta2014}%
  \BibitemOpen
  \bibfield  {author} {\bibinfo {author} {\bibfnamefont {W.~J.}\ \bibnamefont
  {Szlachta}}, \bibinfo {author} {\bibfnamefont {A.~P.}\ \bibnamefont
  {Bart{\'{o}}k}}, \ and\ \bibinfo {author} {\bibfnamefont {G.}~\bibnamefont
  {Cs{\'{a}}nyi}},\ }\href {\doibase 10.1103/PhysRevB.90.104108} {\bibfield
  {journal} {\bibinfo  {journal} {Phys. Rev. B}\ }\textbf {\bibinfo {volume}
  {90}},\ \bibinfo {pages} {104108} (\bibinfo {year} {2014})}\BibitemShut
  {NoStop}%
\bibitem [{\citenamefont {Deringer}\ and\ \citenamefont
  {Cs{\'{a}}nyi}(2017)}]{aC_GAP}%
  \BibitemOpen
  \bibfield  {author} {\bibinfo {author} {\bibfnamefont {V.~L.}\ \bibnamefont
  {Deringer}}\ and\ \bibinfo {author} {\bibfnamefont {G.}~\bibnamefont
  {Cs{\'{a}}nyi}},\ }\href@noop {} {\bibfield  {journal} {\bibinfo  {journal}
  {Phys. Rev. B}\ }\textbf {\bibinfo {volume} {95}},\ \bibinfo {pages} {094203}
  (\bibinfo {year} {2017})}\BibitemShut {NoStop}%
\bibitem [{\citenamefont {Botu}\ and\ \citenamefont
  {Ramprasad}(2015)}]{Botu2015}%
  \BibitemOpen
  \bibfield  {author} {\bibinfo {author} {\bibfnamefont {V.}~\bibnamefont
  {Botu}}\ and\ \bibinfo {author} {\bibfnamefont {R.}~\bibnamefont
  {Ramprasad}},\ }\href {\doibase 10.1103/PhysRevB.92.094306} {\bibfield
  {journal} {\bibinfo  {journal} {Phys. Rev. B}\ }\textbf {\bibinfo {volume}
  {92}},\ \bibinfo {pages} {094306} (\bibinfo {year} {2015})}\BibitemShut
  {NoStop}%
\bibitem [{\citenamefont {Seko}\ \emph {et~al.}(2015)\citenamefont {Seko},
  \citenamefont {Takahashi},\ and\ \citenamefont {Tanaka}}]{Seko2015}%
  \BibitemOpen
  \bibfield  {author} {\bibinfo {author} {\bibfnamefont {A.}~\bibnamefont
  {Seko}}, \bibinfo {author} {\bibfnamefont {A.}~\bibnamefont {Takahashi}}, \
  and\ \bibinfo {author} {\bibfnamefont {I.}~\bibnamefont {Tanaka}},\ }\href
  {\doibase 10.1103/PhysRevB.92.054113} {\bibfield  {journal} {\bibinfo
  {journal} {Phys. Rev. B}\ }\textbf {\bibinfo {volume} {92}},\ \bibinfo
  {pages} {054113} (\bibinfo {year} {2015})}\BibitemShut {NoStop}%
\bibitem{Li2015}
  Z.~Li, J.~R. Kermode, and A.~De Vita,
  Phys. Rev. Lett. {\bf 114}, 096405 (2015). 
\bibitem [{\citenamefont {Shapeev}(2016)}]{Shapeev2016}%
  \BibitemOpen
  \bibfield  {author} {\bibinfo {author} {\bibfnamefont {A.}~\bibnamefont
  {Shapeev}},\ }\href {\doibase 10.1137/15M1054183} {\bibfield  {journal}
  {\bibinfo  {journal} {Multiscale Model. Simul.}\ }\textbf {\bibinfo {volume}
  {14}},\ \bibinfo {pages} {1153} (\bibinfo {year} {2016})}\BibitemShut
  {NoStop}%
\bibitem{Kruglov2017}
  I.~Kruglov, O.~Sergeev, A.~Yanilkin, and A.~R. Oganov,
  Sci. Rep. {\bf 7}, 8512 (2017).
\bibitem{Podryabinkin2017}
  E.~V. Podryabinkin and A.~V. Shapeev,
  Comput. Mater. Sci. {\bf 140}, 171 (2017).
\bibitem{Huan2017}
  T.~D. Huan, R.~Batra, J.~Chapman, S.~Krishnan, L.~Chen, and R.~Ramprasad,
  npj Comput. Mater. {\bf 3}, 37 (2017).
\bibitem [{\citenamefont {Sosso}\ \emph {et~al.}(2013)\citenamefont {Sosso},
  \citenamefont {Miceli}, \citenamefont {Caravati}, \citenamefont {Giberti},
  \citenamefont {Behler},\ and\ \citenamefont {Bernasconi}}]{Sosso2013}%
  \BibitemOpen
  \bibfield  {author} {\bibinfo {author} {\bibfnamefont {G.~C.}\ \bibnamefont
  {Sosso}}, \bibinfo {author} {\bibfnamefont {G.}~\bibnamefont {Miceli}},
  \bibinfo {author} {\bibfnamefont {S.}~\bibnamefont {Caravati}}, \bibinfo
  {author} {\bibfnamefont {F.}~\bibnamefont {Giberti}}, \bibinfo {author}
  {\bibfnamefont {J.}~\bibnamefont {Behler}}, \ and\ \bibinfo {author}
  {\bibfnamefont {M.}~\bibnamefont {Bernasconi}},\ }\href {\doibase
  10.1021/jz402268v} {\bibfield  {journal} {\bibinfo  {journal} {J. Phys. Chem.
  Lett.}\ }\textbf {\bibinfo {volume} {4}},\ \bibinfo {pages} {4241} (\bibinfo
  {year} {2013})}\BibitemShut {NoStop}%
\bibitem [{\citenamefont {Sosso}\ \emph {et~al.}(2015)\citenamefont {Sosso},
  \citenamefont {Salvalaglio}, \citenamefont {Behler}, \citenamefont
  {Bernasconi},\ and\ \citenamefont {Parrinello}}]{Sosso2015a}%
  \BibitemOpen
  \bibfield  {author} {\bibinfo {author} {\bibfnamefont {G.~C.}\ \bibnamefont
  {Sosso}}, \bibinfo {author} {\bibfnamefont {M.}~\bibnamefont {Salvalaglio}},
  \bibinfo {author} {\bibfnamefont {J.}~\bibnamefont {Behler}}, \bibinfo
  {author} {\bibfnamefont {M.}~\bibnamefont {Bernasconi}}, \ and\ \bibinfo
  {author} {\bibfnamefont {M.}~\bibnamefont {Parrinello}},\ }\href {\doibase
  10.1021/acs.jpcc.5b00296} {\bibfield  {journal} {\bibinfo  {journal} {J.
  Phys. Chem. C}\ }\textbf {\bibinfo {volume} {119}},\ \bibinfo {pages} {6428}
  (\bibinfo {year} {2015})}\BibitemShut {NoStop}%
\bibitem{Gabardi2017} S.~Gabardi, E.~Baldi, E.~Bosoni, D.~Campi, S.~Caravati,
  G.~C. Sosso, J.~Behler, M.~Bernasconi, J. Phys. Chem. C {\bf 121},
  23827 (2017).
\bibitem [{\citenamefont {Behler}\ \emph {et~al.}(2008)\citenamefont {Behler},
  \citenamefont {Marton{\'{a}}k}, \citenamefont {Donadio},\ and\ \citenamefont
  {Parrinello}}]{Behler2008}%
  \BibitemOpen
  \bibfield  {author} {\bibinfo {author} {\bibfnamefont {J.}~\bibnamefont
  {Behler}}, \bibinfo {author} {\bibfnamefont {R.}~\bibnamefont
  {Marton{\'{a}}k}}, \bibinfo {author} {\bibfnamefont {D.}~\bibnamefont
  {Donadio}}, \ and\ \bibinfo {author} {\bibfnamefont {M.}~\bibnamefont
  {Parrinello}},\ }\href {\doibase 10.1103/PhysRevLett.100.185501} {\bibfield
  {journal} {\bibinfo  {journal} {Phys. Rev. Lett.}\ }\textbf {\bibinfo
  {volume} {100}},\ \bibinfo {pages} {185501} (\bibinfo {year}
  {2008})}\BibitemShut {NoStop}%
\bibitem [{\citenamefont {Khaliullin}\ \emph {et~al.}(2011)\citenamefont
  {Khaliullin}, \citenamefont {Eshet}, \citenamefont {K{\"{u}}hne},
  \citenamefont {Behler},\ and\ \citenamefont {Parrinello}}]{Khaliullin2011}%
  \BibitemOpen
  \bibfield  {author} {\bibinfo {author} {\bibfnamefont {R.~Z.}\ \bibnamefont
  {Khaliullin}}, \bibinfo {author} {\bibfnamefont {H.}~\bibnamefont {Eshet}},
  \bibinfo {author} {\bibfnamefont {T.~D.}\ \bibnamefont {K{\"{u}}hne}},
  \bibinfo {author} {\bibfnamefont {J.}~\bibnamefont {Behler}}, \ and\ \bibinfo
  {author} {\bibfnamefont {M.}~\bibnamefont {Parrinello}},\ }\href
  {http://dx.doi.org/10.1038/nmat3078
  http://www.nature.com/nmat/journal/v10/n9/abs/nmat3078.html{\#}supplementary-information}
  {\bibfield  {journal} {\bibinfo  {journal} {Nat. Mater.}\ }\textbf {\bibinfo
  {volume} {10}},\ \bibinfo {pages} {693} (\bibinfo {year} {2011})}\BibitemShut
  {NoStop}%
\bibitem [{\citenamefont {Eshet}\ \emph {et~al.}(2012)\citenamefont {Eshet},
  \citenamefont {Khaliullin}, \citenamefont {K{\"{u}}hne}, \citenamefont
  {Behler},\ and\ \citenamefont {Parrinello}}]{Eshet2012}%
  \BibitemOpen
  \bibfield  {author} {\bibinfo {author} {\bibfnamefont {H.}~\bibnamefont
  {Eshet}}, \bibinfo {author} {\bibfnamefont {R.~Z.}\ \bibnamefont
  {Khaliullin}}, \bibinfo {author} {\bibfnamefont {T.~D.}\ \bibnamefont
  {K{\"{u}}hne}}, \bibinfo {author} {\bibfnamefont {J.}~\bibnamefont {Behler}},
  \ and\ \bibinfo {author} {\bibfnamefont {M.}~\bibnamefont {Parrinello}},\
  }\href {\doibase 10.1103/PhysRevLett.108.115701} {\bibfield  {journal}
  {\bibinfo  {journal} {Phys. Rev. Lett.}\ }\textbf {\bibinfo {volume} {108}},\
  \bibinfo {pages} {115701} (\bibinfo {year} {2012})}\BibitemShut {NoStop}%
\bibitem [{\citenamefont {Rupp}(2015)}]{Rupp2015}%
   M.~Rupp, Int. J. Quantum Chem. \textbf{115}, 1058 (2015);
   see also further contributions in the corresponding Special Issue.
\bibitem [{\citenamefont {Behler}(2016)}]{Behler2016}%
   J.~Behler, J. Chem. Phys. \textbf{145}, 170901 (2016).
\bibitem{Laurila2017}
   T.~Laurila, S.~Sainio, and M.~A. Caro,
   Prog. Mater. Sci. {\bf 88}, 499 (2017).
\bibitem [{\citenamefont {Deringer}\ \emph {et~al.}(2017)\citenamefont
  {Deringer}, \citenamefont {Cs{\'{a}}nyi},\ and\ \citenamefont
  {Proserpio}}]{C_AIRSS}%
  \BibitemOpen
  \bibfield  {author} {\bibinfo {author} {\bibfnamefont {V.~L.}\ \bibnamefont
  {Deringer}}, \bibinfo {author} {\bibfnamefont {G.}~\bibnamefont
  {Cs{\'{a}}nyi}}, \ and\ \bibinfo {author} {\bibfnamefont {D.~M.}\
  \bibnamefont {Proserpio}},\ }\href@noop {} {\bibfield  {journal} {\bibinfo
  {journal} {ChemPhysChem}\ }\textbf {\bibinfo {volume} {18}},\ \bibinfo
  {pages} {873} (\bibinfo {year} {2017})}\BibitemShut {NoStop}%
\bibitem{Mahoney2009}
  M.~Mahoney and P.~Drineas, Proc. Natl. Acad. Sci. U. S. A. {\bf 106},
  697 (2009).
\bibitem [{\citenamefont {Bart{\'{o}}k}\ and\ \citenamefont
  {Cs{\'{a}}nyi}(2015)}]{Bartok2015}%
  \BibitemOpen
  \bibfield  {author} {\bibinfo {author} {\bibfnamefont {A.~P.}\ \bibnamefont
  {Bart{\'{o}}k}}\ and\ \bibinfo {author} {\bibfnamefont {G.}~\bibnamefont
  {Cs{\'{a}}nyi}},\ }\href@noop {} {\bibfield  {journal} {\bibinfo  {journal}
  {Int. J. Quantum Chem.}\ }\textbf {\bibinfo {volume} {115}},\ \bibinfo
  {pages} {1051} (\bibinfo {year} {2015})}\BibitemShut {NoStop}%
\bibitem{Li2017a}
  W.~Li, Y.~Ando, and S.~Wantanabe,
  J.\ Phys.\ Soc.\ Jpn. {\bf 86}, 104004 (2017).
\bibitem [{\citenamefont {Bart{\'{o}}k}\ \emph {et~al.}(2013)\citenamefont
  {Bart{\'{o}}k}, \citenamefont {Kondor},\ and\ \citenamefont
  {Cs{\'{a}}nyi}}]{Bartok2013}%
  \BibitemOpen
  \bibfield  {author} {\bibinfo {author} {\bibfnamefont {A.~P.}\ \bibnamefont
  {Bart{\'{o}}k}}, \bibinfo {author} {\bibfnamefont {R.}~\bibnamefont
  {Kondor}}, \ and\ \bibinfo {author} {\bibfnamefont {G.}~\bibnamefont
  {Cs{\'{a}}nyi}},\ }\href {\doibase 10.1103/PhysRevB.87.184115} {\bibfield
  {journal} {\bibinfo  {journal} {Phys. Rev. B}\ }\textbf {\bibinfo {volume}
  {87}},\ \bibinfo {pages} {184115} (\bibinfo {year} {2013})}\BibitemShut
  {NoStop}%
\bibitem [{\citenamefont {Cliffe}\ \emph {et~al.}(2017)\citenamefont {Cliffe},
  \citenamefont {Bart{\'{o}}k}, \citenamefont {Kerber}, \citenamefont {Grey},
  \citenamefont {Cs{\'{a}}nyi},\ and\ \citenamefont {Goodwin}}]{Cliffe2017}%
  \BibitemOpen
  \bibfield  {author} {\bibinfo {author} {\bibfnamefont {M.~J.}\ \bibnamefont
  {Cliffe}}, \bibinfo {author} {\bibfnamefont {A.~P.}\ \bibnamefont
  {Bart{\'{o}}k}}, \bibinfo {author} {\bibfnamefont {R.~N.}\ \bibnamefont
  {Kerber}}, \bibinfo {author} {\bibfnamefont {C.~P.}\ \bibnamefont {Grey}},
  \bibinfo {author} {\bibfnamefont {G.}~\bibnamefont {Cs{\'{a}}nyi}}, \ and\
  \bibinfo {author} {\bibfnamefont {A.~L.}\ \bibnamefont {Goodwin}},\
  }\href@noop {} {\bibfield  {journal} {\bibinfo  {journal} {Phys. Rev. B}\ }\textbf {\bibinfo {volume} {95}},\ \bibinfo
  {pages} {224108} (\bibinfo {year} {2017})}\BibitemShut {NoStop}%
\bibitem [{\citenamefont {De}\ \emph {et~al.}(2016)\citenamefont {De},
  \citenamefont {Bart{\'{o}}k}, \citenamefont {Cs{\'{a}}nyi},\ and\
  \citenamefont {Ceriotti}}]{De2016}%
  \BibitemOpen
  \bibfield  {author} {\bibinfo {author} {\bibfnamefont {S.}~\bibnamefont
  {De}}, \bibinfo {author} {\bibfnamefont {A.~P.}\ \bibnamefont
  {Bart{\'{o}}k}}, \bibinfo {author} {\bibfnamefont {G.}~\bibnamefont
  {Cs{\'{a}}nyi}}, \ and\ \bibinfo {author} {\bibfnamefont {M.}~\bibnamefont
  {Ceriotti}},\ }\href {\doibase 10.1039/C6CP00415F} {\bibfield  {journal}
  {\bibinfo  {journal} {Phys. Chem. Chem. Phys.}\ }\textbf {\bibinfo {volume}
  {18}},\ \bibinfo {pages} {13754} (\bibinfo {year} {2016})}\BibitemShut
  {NoStop}%
\bibitem{Bartok2013a}
 A.~P. Bart\'o{}k, M.~J. Gillan, F.~R. Manby, and G.~Cs\'a{}nyi,
 Phys. Rev. B {\bf 88}, 054104 (2013).
\bibitem{Gobre2013}
 V.~V. Gobre, A. Tkatchenko, Nat. Commun. {\bf 4}, 2341 (2013).
\bibitem [{\citenamefont {Ganesh}\ \emph {et~al.}(2014)\citenamefont {Ganesh},
  \citenamefont {Kim}, \citenamefont {Park}, \citenamefont {Yoon},
  \citenamefont {Reboredo},\ and\ \citenamefont {Kent}}]{Ganesh2014}%
  \BibitemOpen
  \bibfield  {author} {\bibinfo {author} {\bibfnamefont {P.}~\bibnamefont
  {Ganesh}}, \bibinfo {author} {\bibfnamefont {J.}~\bibnamefont {Kim}},
  \bibinfo {author} {\bibfnamefont {C.}~\bibnamefont {Park}}, \bibinfo {author}
  {\bibfnamefont {M.}~\bibnamefont {Yoon}}, \bibinfo {author} {\bibfnamefont
  {F.~A.}\ \bibnamefont {Reboredo}}, \ and\ \bibinfo {author} {\bibfnamefont
  {P.~R.~C.}\ \bibnamefont {Kent}},\ }\href {\doibase 10.1021/ct500617z}
  {\bibfield  {journal} {\bibinfo  {journal} {J. Chem. Theory Comput.}\
  }\textbf {\bibinfo {volume} {10}},\ \bibinfo {pages} {5318} (\bibinfo {year}
  {2014})}\BibitemShut {NoStop}%
\bibitem [{\citenamefont {Clark}\ \emph {et~al.}(2005)\citenamefont {Clark},
  \citenamefont {Segall}, \citenamefont {Pickard}, \citenamefont {Hasnip},
  \citenamefont {Probert}, \citenamefont {Refson},\ and\ \citenamefont
  {Payne}}]{CASTEP}%
  \BibitemOpen
  \bibfield  {author} {\bibinfo {author} {\bibfnamefont {S.~J.}\ \bibnamefont
  {Clark}}, \bibinfo {author} {\bibfnamefont {M.~D.}\ \bibnamefont {Segall}},
  \bibinfo {author} {\bibfnamefont {C.~J.}\ \bibnamefont {Pickard}}, \bibinfo
  {author} {\bibfnamefont {P.~J.}\ \bibnamefont {Hasnip}}, \bibinfo {author}
  {\bibfnamefont {M.~J.}\ \bibnamefont {Probert}}, \bibinfo {author}
  {\bibfnamefont {K.}~\bibnamefont {Refson}}, \ and\ \bibinfo {author}
  {\bibfnamefont {M.~C.}\ \bibnamefont {Payne}},\ }\href@noop {} {\bibfield
  {journal} {\bibinfo  {journal} {Z. Kristallogr.}\ }\textbf {\bibinfo {volume}
  {220}},\ \bibinfo {pages} {567} (\bibinfo {year} {2005})}\BibitemShut
  {NoStop}%
\bibitem [{\citenamefont {Francis}\ and\ \citenamefont {Payne}(1990)}]{FBSC}%
  \BibitemOpen
  \bibfield  {author} {\bibinfo {author} {\bibfnamefont {G.~P.}\ \bibnamefont
  {Francis}}\ and\ \bibinfo {author} {\bibfnamefont {M.~C.}\ \bibnamefont
  {Payne}},\ }\href@noop {} {\bibfield  {journal} {\bibinfo  {journal} {J.
  Phys.: Condens. Matter}\ }\textbf {\bibinfo {volume} {2}},\ \bibinfo {pages}
  {4395} (\bibinfo {year} {1990})}\BibitemShut {NoStop}%
\bibitem [{\citenamefont {Kresse}\ and\ \citenamefont
  {Hafner}(1993)}]{Kresse1993}%
  \BibitemOpen
  \bibfield  {author} {\bibinfo {author} {\bibfnamefont {G.}~\bibnamefont
  {Kresse}}\ and\ \bibinfo {author} {\bibfnamefont {J.}~\bibnamefont
  {Hafner}},\ }\href@noop {} {\bibfield  {journal} {\bibinfo  {journal} {Phys.
  Rev. B}\ }\textbf {\bibinfo {volume} {47}},\ \bibinfo {pages} {558} (\bibinfo
  {year} {1993})}\BibitemShut {NoStop}%
\bibitem [{\citenamefont {Kresse}\ and\ \citenamefont
  {Furthm{\"{u}}ller}(1996)}]{Kresse1996a}%
  \BibitemOpen
  \bibfield  {author} {\bibinfo {author} {\bibfnamefont {G.}~\bibnamefont
  {Kresse}}\ and\ \bibinfo {author} {\bibfnamefont {J.}~\bibnamefont
  {Furthm{\"{u}}ller}},\ }\href@noop {} {\bibfield  {journal} {\bibinfo
  {journal} {Phys. Rev. B}\ }\textbf {\bibinfo {volume} {54}},\ \bibinfo
  {pages} {11169} (\bibinfo {year} {1996})}\BibitemShut {NoStop}%
\bibitem [{\citenamefont {Kresse}\ and\ \citenamefont
  {Joubert}(1999)}]{Kresse1999}%
  \BibitemOpen
  \bibfield  {author} {\bibinfo {author} {\bibfnamefont {G.}~\bibnamefont
  {Kresse}}\ and\ \bibinfo {author} {\bibfnamefont {D.}~\bibnamefont
  {Joubert}},\ }\href@noop {} {\bibfield  {journal} {\bibinfo  {journal} {Phys.
  Rev. B}\ }\textbf {\bibinfo {volume} {59}},\ \bibinfo {pages} {1758}
  (\bibinfo {year} {1999})}\BibitemShut {NoStop}%
\bibitem [{\citenamefont {Bl{\"{o}}chl}(1994)}]{Blochl1994}%
  \BibitemOpen
  \bibfield  {author} {\bibinfo {author} {\bibfnamefont {P.~E.}\ \bibnamefont
  {Bl{\"{o}}chl}},\ }\href@noop {} {\bibfield  {journal} {\bibinfo  {journal}
  {Phys. Rev. B}\ }\textbf {\bibinfo {volume} {50}},\ \bibinfo {pages} {17953}
  (\bibinfo {year} {1994})}\BibitemShut {NoStop}%
\bibitem{Liu1996}
  P. Liu and H. Wu, Sol. State Ionics {\bf 92}, 91 (1996).
\bibitem [{\citenamefont {Meunier}\ \emph {et~al.}(2002)\citenamefont
  {Meunier}, \citenamefont {Kephart}, \citenamefont {Roland},\ and\
  \citenamefont {Bernholc}}]{Meunier2002}%
  \BibitemOpen
  \bibfield  {author} {\bibinfo {author} {\bibfnamefont {V.}~\bibnamefont
  {Meunier}}, \bibinfo {author} {\bibfnamefont {J.}~\bibnamefont {Kephart}},
  \bibinfo {author} {\bibfnamefont {C.}~\bibnamefont {Roland}}, \ and\ \bibinfo
  {author} {\bibfnamefont {J.}~\bibnamefont {Bernholc}},\ }\href
  {https://link.aps.org/doi/10.1103/PhysRevLett.88.075506} {\bibfield
  {journal} {\bibinfo  {journal} {Phys. Rev. Lett.}\ }\textbf {\bibinfo
  {volume} {88}},\ \bibinfo {pages} {075506} (\bibinfo {year}
  {2002})}\BibitemShut {NoStop}%
\bibitem [{Note1()}]{Note1}%
  \BibitemOpen
  \bibinfo {note} {This structure was generated by annealing a disordered
  amorphous carbon structure using GAP, inducing graphitization following the
  general ideas in R.\ C.\ Powles, N.\ A.\ Marks, and D.\ W.\ M.\ Lau, Phys.\
  Rev.\ B {\protect \bf 79}, 075430 (2009). A more detailed account of this
  will be published elsewhere.}\BibitemShut {Stop}%
\end{thebibliography}
\end{document}